\documentclass[sn-mathphys,Numbered]{sn-jnl}
\usepackage[utf8]{inputenc}
\usepackage{graphicx}
\usepackage{amsmath}
\usepackage{amssymb}
\usepackage{amsthm}
\usepackage{amsfonts}
\usepackage{caption}
\usepackage{subfig}
\usepackage{makeidx}
\usepackage{bm}
\usepackage{soul}
\usepackage{ascmac}
\usepackage{color}
\usepackage{listings}
\usepackage{verbatim}
\usepackage{nicefrac} 
\newcommand{\bra}[1]{\langle #1 |}
\newcommand{\ket}[1]{| #1 \rangle}

\usepackage{multirow}%
\usepackage{mathrsfs}%
\usepackage[title]{appendix}%
\usepackage{xcolor}%
\usepackage{textcomp}%
\usepackage{manyfoot}%
\usepackage{booktabs}%
\usepackage{algorithm}%
\usepackage{algorithmicx}%
\usepackage{algpseudocode}%
\usepackage{siunitx}

\begin{document}
\title{On-Premises Superconducting Quantum Computer for Education and Research}

\author[1]{\fnm{Jami} \sur{ R\"onkk\"o}}
\author[1]{\fnm{Olli} \sur{Ahonen}}
\author[1]{\fnm{Ville} \sur{Bergholm}}
\author[2]{\fnm{Alessio} \sur{Calzona}}
\author[2]{\fnm{Attila} \sur{Geresdi}}
\author[1]{\fnm{Hermanni} \sur{Heimonen}}
\author[1]{\fnm{Johannes} \sur{Heinsoo}}
\author[1]{\fnm{Vladimir} \sur{Milchakov}}
\author[2]{\fnm{Stefan} \sur{Pogorzalek}}
\author[1]{\fnm{Matthew} \sur{Sarsby}}
\author[1]{\fnm{Mykhailo} \sur{Savytskyi}}
\author[2]{\fnm{Stefan} \sur{Seegerer}}
\author[2]{\fnm{Fedor} \sur{\v{S}imkovic IV}}
\author[2]{\fnm{P.V.} \sur{Sriluckshmy}}
\author[1]{\fnm{Panu T.} \sur{Vesanen}}

\author*[1]{\fnm{Mikio} \sur{Nakahara}}\email{mikio.nakahara@meetiqm.com}

\affil[1]{
\orgname{IQM Quantum Computers}, \orgaddress{\street{Keilaranta 19}, \city{Espoo}, \postcode{02150}, 
\country{Finland}}}

\affil[2]{
\orgname{IQM Quantum Computers}, \orgaddress{\street{Georg-Brauchle-Ring 23-25}, \city{Munich}, \postcode{80992}, 
\country{Germany}}}

\date{\today}

\abstract{
\par With a growing interest in quantum technology globally, there is an increasing need for accessing relevant physical systems for education and research. In this paper we introduce a commercially available on-site quantum computer utilizing superconducting technology, offering insights into its fundamental hardware and software components. We show how this system can be used in education to teach quantum concepts and deepen understanding of quantum theory and quantum computing. It offers learning opportunities for future talent and contributes to technological progress. Additionally, we demonstrate its use in research by replicating some notable recent achievements. 
}


\maketitle

\section{Introduction}
\label{sec:introduction}

Quantum computing is a promising technology which is expected to efficiently solve certain classes of problems that are challenging for classical computers in terms of computational time and/or hardware resources. The term \emph{quantum advantage} is coined for the demonstration of this algorithmic speed-up on quantum hardware. Several quantum algorithms have been devised to demonstrate this, for example prime number factorization of large integers~\cite{bib:shor} with an exponential speed-up compared to its best known classical counterpart. A similar speed-up exists when simulating the chemical and physical properties of molecules and the dynamics of fundamental physical models \cite{bib:Daley2022}. We should note that most of these algorithms assume an error-free quantum hardware with a number of quantum bits or qubits beyond the reach of current technology. With the inherent presence of the loss of quantum information in any physical system, a fault-tolerant quantum computer \cite{bib:shor_FTQC} would employ a built-in quantum error correction, where the number of error-free logical qubits is less than the error-prone and noisy physical qubits.

However, even in the absence of error-correction, noisy intermediate-scale quantum (NISQ) computers \cite{bib:RevModPhys.94.015004} are thought to exhibit quantum advantage over classical high-performance computers (HPC) in the range of 100 to 1000 qubits, depending on the quality of the quantum hardware and the connectivity between the qubits. Among many physical platforms, superconducting quantum hardware is well-suited for scaling the number of qubits and improving their fidelity while maintaining connectivity and thus becomes a preferred technology in the NISQ era with roadmaps towards fault tolerance \cite{bib:10.1126/science.1231930}.

Here, we introduce the IQM Spark\textsuperscript{TM}~\cite{bib:spark} prototype\footnote{This research paper is based on a prototype system of a commercial IQM Spark\textsuperscript{TM} and its system specifications and performance may therefore differ from the commercial IQM Spark\textsuperscript{TM}. The authors will not make any statements relating to the commercial IQM Spark\textsuperscript{TM} or give any guarantees relating to the system specification and performance of the commercial IQM Spark\textsuperscript{TM}.}, a 5-qubit superconducting quantum computer designed and developed to enable a low-barrier access to both its hardware and software components. The hardware is self-contained with a packaged superconducting quantum processing unit (QPU), a dilution refrigerator, control electronics, while the software components allow for both a direct manipulation of the qubits by microwave pulses or to run small scale quantum algorithms composed of quantum gates~\cite{bib:spark}. 
As we will demonstrate with different use cases, this system can be harnessed for a range of educational activities from teaching the concepts of superconducting quantum hardware to developing an understanding of quantum error mitigation and performing experiments from different fields of research.

The rest of the paper is organized as follows. In Section~\ref{sec:hardware}, we introduce the hardware, in which, after the overview, the basics of the transmon qubits, tunable couplers, traveling wave parametric amplifiers, dilution refrigerators, and control electronics are explained. In Section~\ref{sec:software}, we introduce the software necessary to operate the quantum computer. The next two sections are dedicated to applications of a quantum computer to education and research. Some applications are available only for an on-premises quantum computer.
In Section~\ref{sec:education} we introduce use cases for educational purposes, namely
\begin{itemize}
    \item calibration,
    \item benchmarking
    \item visualization of pulses with oscilloscope
       \item error mitigation, and finally
    \item execution of simple quantum algorithms.
\end{itemize}
In Section~\ref{sec:research}, we reproduce some research results which appeared recently in scientific journals, namely
\begin{itemize}
    \item simulation of neutrino oscillation,
    \item estimation of Jones polynomials, and
    \item an introduction into embedding techniques for quantum chemistry.
\end{itemize}
Section~\ref{sec:discussion} is devoted to summary and discussion.

\subsection*{Notation}

Throughout this paper, $I_k$ stands for the $k$-dimensional unit matrix. Pauli matrices are denoted by $X, Y$ and~$Z$, with an optional subscript to denote the qubit they apply to:
$$
P_i = I_2 \otimes \ldots \otimes I_2 \otimes P \otimes I_2 
\otimes \ldots \otimes I_2, \quad P \in\{X,Y,Z\},
$$
where $P$ is in the $i$th position. $X_1 Z_3=X \otimes I_2 \otimes Z$, for example. We use the common convention for qubit ordering, namely the top qubit in a quantum circuit is the first qubit, which is opposite to the Qiskit convention~\cite{bib:Qiskit}. We assign qubit numbers 1 to 5 as shown in Fig.~1 although Qiskit assigns 0 to 4. Which convention is employed should be obvious from the context.

\subsection*{Estimating expectation values of observables}

A quantum computer estimates the probabilities of the $Z$-basis states of the measured qubits by repeating the same multiqubit $Z$-basis measurement many times, and computing the relative frequencies of the outcomes. The number of repetitions is called ``shots''. For example, for a single-qubit state $\ket{\psi} =a\ket{0}+b\ket{1}$ we may estimate the probabilities $p_0=|a|^2$ and $p_1=|b|^2$. Using these values we may estimate
the expectation value of $Z_i$ as
$\bra{\psi} Z_i \ket{\psi}=p_0 - p_1$.
By using the identities
\begin{equation}
X = H\,Z\,H \quad \text{and}
\quad Y = S\,H\, Z \,H\,S^\dagger,
\end{equation}
we may estimate the expectation value of $X$ or $Y$ by first rotating the state $\ket{\psi}$ by $H$ or $H\,S^\dagger$, respectively, and then estimate the expectation value of~$Z$. For execution on our hardware, all circuits are transpiled into single-qubit $R(\theta, \phi)=\exp\left[-i\theta\left(\cos\phi\, X + \sin \phi \,Y\right)/2 \right]$~gates and two-qubit $CZ$~gates. In this process the extra gates used for estimating $X$- or $Y$-expectation values are often merged with adjacent single-qubit gates in the circuit.

\section{Hardware}
\label{sec:hardware}

\par Our superconducting quantum computer is a full-stack system consisting of a 5-qubit superconducting QPU, dilution refrigerator, optimized cryogenic microwave and DC lines, control electronics, and appropriate classical computing hardware to run the control software. The details of our quantum computer used in the experiments of this paper are described in the following subsections.

\subsection{Quantum processing unit}

    \subsubsection{Overview}
            
        \begin{figure}
        \centering%
        \includegraphics[width=0.8\textwidth]{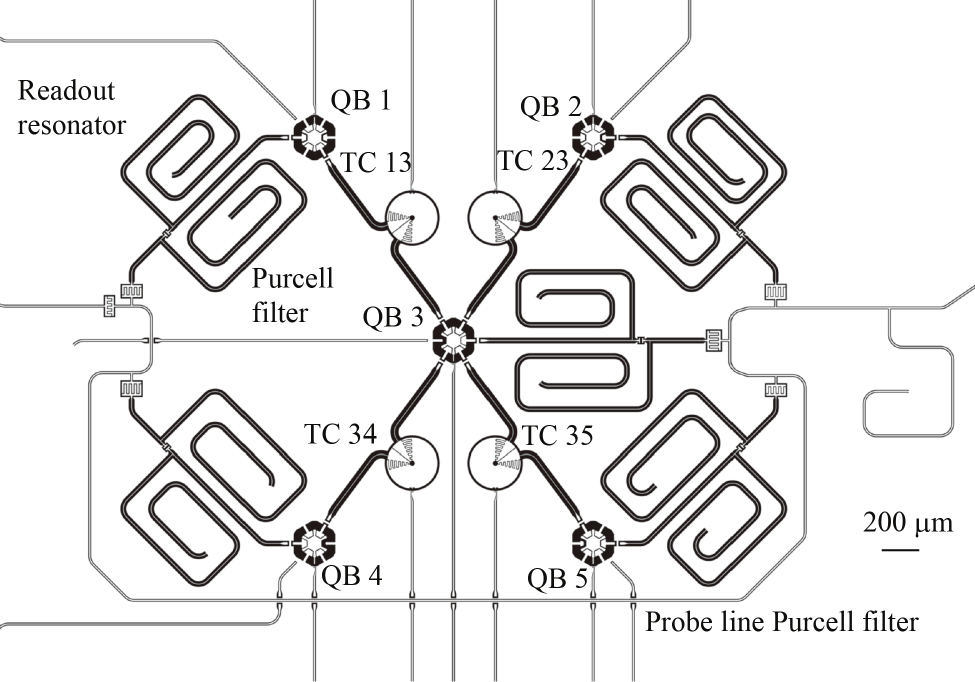}
        \caption{Design of a 5-qubit superconducting quantum processing unit employed in this paper, showing 5 qubits (QB) connected by 4 tunable couplers (TC). Black, apart from explanatory text, indicates areas where superconducting film is etched exposing the substrate.
        }
        \label{QPU_topology}
        \end{figure}
        
        \par The core of any quantum computing system is the quantum processing unit (QPU) comprising of  qubits, qubit-couplers and control and readout lines.
        The QPU of our quantum computer features five data qubits with a single central qubit connected to the four peripheral qubit through tunable couplers~\cite{tunable_coupler_mit, fabian_coupler} in star topology as depicted in Fig.~\ref{QPU_topology}.
        The qubits and tunable couplers are described below in more detail.
        \par The chip layout is drawn using KQCircuits~\cite{kqc}. It is a free open-source add-on for KLayout~\cite{klayout}, a widely used open-source operating system independent layout viewer and editor for integrated circuits. KQCircuits adds the necessary functions to KLayout to programmatically, or using graphical user interface, draw superconducting circuits and export the designs for fabrication or for standard microwave simulation software. KQCircuits also exports netlists, which enable SPICE-like quasi-lumped-element simulations for fast validation and optimization of geometrical parameters to achieve target coupling strengths between circuit elements discussed below.

    \subsubsection{Qubit type}

        \par The states of computational qubits in our QPU are physically stored in non-linear oscillators referred to as \textit{transmon qubits}. Transmon is a modified charge qubit, where a Josephson junction or a Superconducting QUantum Interface Device (SQUID) is shunted by a large capacitor in a way that Josephson energy exceeds capacitor energy by a factor of few tens~\cite{Transmon_Koch}.
        The transmon is prevalent qubit type in superconducting quantum computation due to its stability against charge and flux noise, and simplicity of operation.
        In our qubit circuit, sometimes referred to as a \textit{grounded transmon}~\cite{PhysRevLett_111_080502}, the qubit capacitor is formed by a thin metal film island separated from the coplanar ground plane by a gap where metal has been etched and underling dielectric is exposed. In addition to the mutual capacitance, the central island is connected to the ground via the SQUID consisting of parallel-connected Josephson junctions. 
        
        The shape of the qubit charge island has six-fold rotational symmetry. Each sector features a capacitor island. Each of the islands has its own size, which allows the coupling capacitance to be individually tuned to achieve target coupling to the neighboring qubits, qubit state readout resonator and extra shunt to the ground for precise targeting of total shunt capacitance. In between two coupling islands, there is a narrow strip of charge island to reduce coupling between the neighboring couplers.
    
    \subsubsection{Qubit control}
        \par Each of the qubits is individually addressed by two control lines. Control lines are implemented as coplanar waveguides. 
        
        \par The center conductor of the \textit{flux line} is shorted to the ground in the vicinity of the qubit SQUID creating an effective mutual inductance between the center conductor and the SQUID loop. By applying electrical current through the flux line, magnetic flux created through the SQUID loop creates a phase bias across the Josephson junctions, reducing the effective Josephson energy of the SQUID and hence the qubit frequency.
        At the maximum qubit frequency, the frequency is insensitive, to the first order, to external flux maximizing the coherence time of the qubit and is referred to as a \textit{sweetspot}. Qubit frequency changes are used to find overall optimal operation frequencies, change dispersive coupling rates to the other elements, and implement physical $Z$ and $CZ$ gates, more on two-qubit gates below.

        \par For \textit{drive lines}, the center conductor is left open circuited and due to proximity has mutual capacitance to the charge island of the qubit. By applying microwave signals through the drive line, the qubit state can be driven between internal states with the energy difference corresponding the signal frequency. 
        The capacitance value is carefully chosen to ensure sufficiently low qubit coupling to the $50\, \Omega$ environment to minimize Purcell losses ~\cite{Purcell_Schoelkopf} whilst keeping the coupling to the target qubit strong compared to the spectator qubits. Resonant qubit drive is used to implement $X$, $Y$ or any $R(\theta, \phi)$ gate. By choosing corresponding drive frequency, gates between higher energy levels of the qubit can also be achieved giving access to the Hilbert space of larger dimension.

    \subsubsection{Readout}
        
        To infer the state of a superconducting qubit, so-called dispersive readout is employed~\cite{blais2004, wallraff2005approaching}.
        This is a widely used method which employs transverse coupling between the qubit and resonator based on a dipole-dipole interaction. 
        Due to the transverse coupling term in the Jaynes-Cummings Hamiltonian~\cite{JC_hamiltonian}, a qubit-state-dependent frequency shift of the resonator, known as the \textit{dispersive shift}, is observed.
        Every qubit has a dedicated readout resonator connected to it and each readout resonator has a different resonance frequency.
        
        \par To suppress the Purcell decay rate of the qubits through the readout resonators, the resonator is not directly coupled to the \SI{50}{\ohm} environment. Instead,  each readout resonator couples to an individual \textit{Purcell filter} - a bandpass filter which reduces the transmission at the qubit frequency.
        
\begin{figure}
        \centering%
        \includegraphics[width=0.5\textwidth]{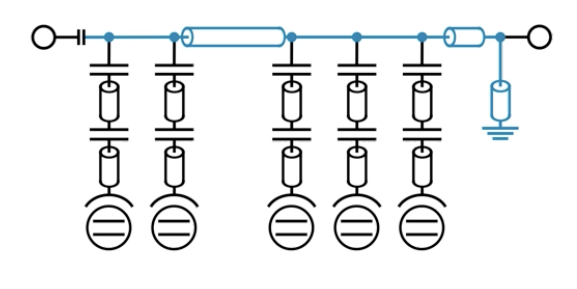}
        \caption{Quasi lumped element circuit diagram of readout circuit, including readout- and Purcell resonators connected to probe line, which consists of a distributed Purcell filter. Qubits are depicted as circles with two horizontal lines.
        }
        \label{readout_circuit_scheme}
        \end{figure}

        As shown in Fig.~\ref{readout_circuit_scheme}, the Purcell filters are in turn all coupled to a common \textit{probe line}. The state of the qubit registry is inferred by probing  the transmission of the probe line with a frequency comb and comparing the phase and amplitude of the transmitted signal at the frequency of each readout resonator individually to a set threshold. The readout cross-talk is reduced thanks to the individual Purcell filters~\cite{Heinsoo2018}.
        The amount of coupling and the frequency detuning between the elements is optimised to balance readout speed and Purcell relaxation rate. The input capacitor and the shunt at the output of the probe line forms another resonator, where the total length defines the frequency and the location of the output tap defines coupling strength to the output port, see the right side of Fig.~\ref{readout_circuit_scheme}.

        The coupling strengths and detunings are chosen such that the Purcell effect would not limit the intrinsic $T_1$.
        
    \subsubsection{Tunable couplers}
        \par
\begin{figure}
        \centering
        \includegraphics[width=0.8\textwidth]{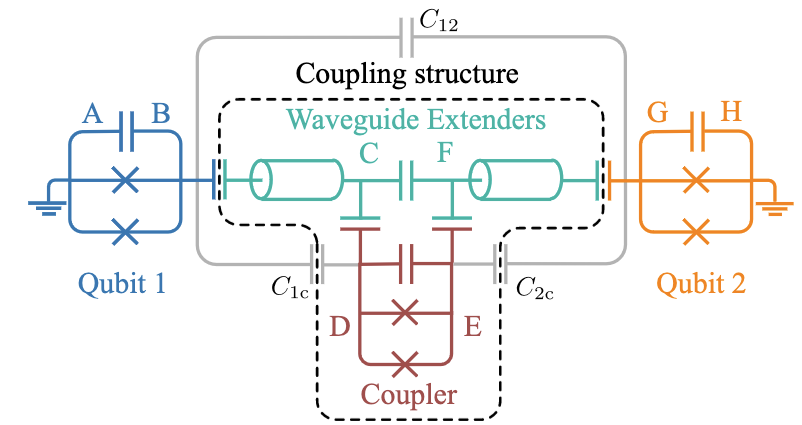}
        \caption{A quasi-lumped element circuit diagram of two transmon qubits (blue and orange) coupled by a tunable coupling structure consisting of waveguide extenders (turquoise) and a floating coupler qubit (red)~\cite{fabian_coupler}. 
        Electrical nodes are marked with capital letters. 
        Grey color elements represent the effective couplings implemented by the waveguide extenders.
        }
        \label{coupler_scheme}
        \end{figure}

 Tunable couplers are utilized in order to perform two-qubit gates between the above-mentioned qubits. 
        Tunable couplers are circuit components based on transmon qubits, which enable us to perform two-qubits gates with state of the art fidelities above 99\%~\cite{fabian_coupler}. 
        The main benefit of using tunable couplers is the possibility to compensate the native $ZZ$-interaction between qubits which enables high fidelity identity gates~\cite{tunable_coupler_mit}.
        In our design, see Fig.~\ref{coupler_scheme}, the interaction between tunable coupler and qubits is mediated by waveguide extenders~\cite{fabian_coupler}.
        This feature allows us to place a significant distance between the qubits to avoid inter-qubit cross-talk, while keeping the switchable $ZZ$-coupling large enough and, in larger devices, fit the readout resonators into the qubit lattice unit cell. 
        
        \par By applying an external magnetic flux into the SQUID loop of a tunable coupler, one can change a coupler frequency and thus the effective amount of the $ZZ$-interaction $g_{zz}$ between relevant pair of qubits.
        The effective value of the $g_{zz}$ can be changed in a wide range including both positive and negative values.
        Consequently, there exists a point where this interaction is equal to zero. 
        This point is used while the QPU is idling so that all qubit pairs have negligible interaction and all the native couplings between pairs of qubits are compensated.
        To perform a two-qubit gate between neighboring qubits, the coupler can be flux-tuned by a square-like baseband pulse to change its frequency causing an interaction between qubits for a certain amount of time, see also subsection \ref{control_electronics_subsection}.
        
        \par By choosing the detuning between qubits during a gate operation, one can implement $CZ$, $i$SWAP gates or any general fermionic simulation gate~\cite{foxen2020} with the same hardware.

\subsection{QPU Package}

    \par The QPU is mounted inside a carrier for handling, shielding, mounting, and signal connection purposes. 
    The shape of the carrier is optimized to make all standing wave modes be far-detuned from operational frequencies.
    The QPU is wire-bonded to a printed circuit board (PCB) with coplanar waveguides that is also attached to the carrier. 
    The PCB serves to transmit signals between QPU launchpads and external microwave connections. 
    The carrier is manufactured out of high conductivity copper for improved thermalisation and reducing potential interactions with impurities in the metal. This carrier has a gold plated finish for better thermal contact between mating surfaces, to reduce losses of signals in the exposed transmission lines, and to ensure high quality factors of all mentioned standing wave modes. The particular gold plating process used is non-magnetic to maintain a clean magnetic field environment. 
    The sample carrier securely holds the QPU, protects the chip from stray radiation, minimizes microwave interference and cross-talk between the  signals.

    To thermally attach the QPU carrier to the refrigerator, it is mounted to a copper cold finger, which is situated inside a multiple layered magnetic shielding assembly.
    These shields are required for minimizing the interaction with external magnetic fields and suppress to environmental radiation.

\subsection{Refrigerator}

    \par The cold finger with the chip carrier is attached to the mixing chamber plate of a commercial Bluefors\textsuperscript{TM} refrigerator in order to cool the QPU to the operating temperature of a few tens of mK.
    These machines use a pulse tube refrigerator to cool the first stages to a few Kelvin and then use a dilution refrigerator to reach the base temperature.
    The experiments presented in this paper were performed with components attached to the experimental stage with a temperature of 30\,mK or cooler.

\subsection{Signal inputs and outputs}

    \par The microwave signals for the qubit drive, tunable coupler flux, parametric amplifier pump, and readout probe are routed from room temperature to the QPU by coaxial SCuNi wires that include appropriate attenuation cascade at the different temperature stages as well as low-pass filtering at the base temperature. DC signals for qubit flux are routed using twisted pair wiring with low-pass filtering at room temperature, the 3\,K stage, and at the base temperature stage. 
    \par The readout response is amplified by a Traveling Wave Parametric Amplifier (TWPA), routed via appropriate isolation upwards by superconducting coaxial NbTi wiring to a High-Electron-Mobility Transistor (HEMT) amplifier operating at a nominal 3\,K temperature. After the HEMT the signal is carried by silver plated copper-nickel coaxial wiring to reach the top plate of the cryostat. 
    
    \subsubsection{Traveling wave parametric amplifier}

        Qubit readout relies on readout pulses with as little energy as tens of microwave photons. To detect such weak signals, quantum limited amplifiers are used, where amount of added noise is limited by quantum mechanics~\cite{caves1982quantum}. The first amplifier in our readout chain is a 
        TWPA. Due to high gain and bandwidth, it enables frequency multiplxed readout of all qubits within \SI{100}{\nano\second} and readout fidelity limited only by qubit decay time.~\cite{swiadek2023}
        
\begin{figure}[tb]
        \centering
        \includegraphics[width=0.7\textwidth]{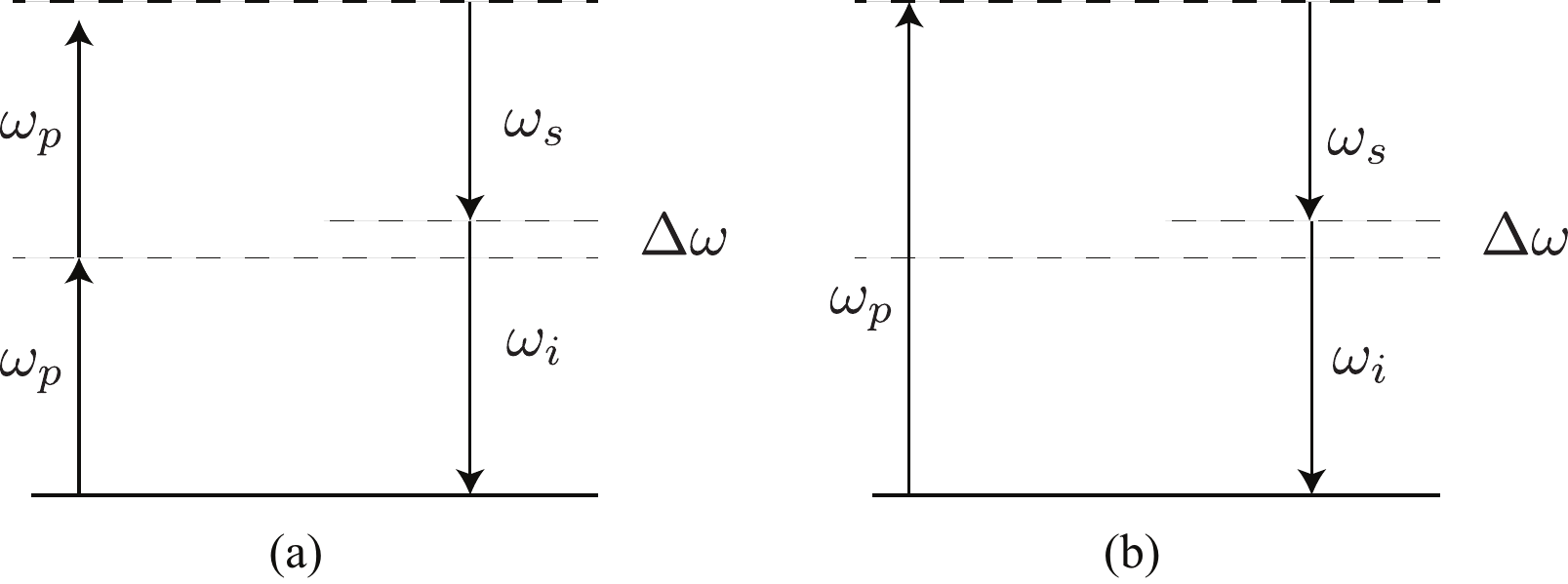}
        \caption{Energy level diagrams for (a) 4WM and (b) 3WM processes. $\omega_p$, $\omega_s$ and $\omega_i$ represent the angular frequencies of pump, signal and idler photons, respectively. $\Delta\omega$ represents an frequency detuning from the degenerate mode of amplification.}
        \label{wave_mixing}
        \end{figure}

        Parametric amplifiers include nonlinear media, where propagating weak and strong tones exchange energy.
        In TWPAs, the nonlinear media consists of long series of Josephson junctions \cite{frattini20173} forming an analogue of an optical Kerr medium.
        The Josephson potential being an even function of the superconducting phase difference $\phi$, the provided nonlinearity to the lowest order is of the form $\phi^4$ which enables four-wave mixing (4WM). Here two  photons from a strong \textit{pump} tone generate one photon in phase with the weak input \textit{signal} and another \textit{idler} photon~\cite{macklin2015near}. Pump, signal and idler tone frequencies are related by the conservation of energy~\cite{caves1982quantum}, see Fig.~\ref{wave_mixing}.
        
        In the presence of an external flux, the lowest order of non-linearity becomes of the form $\phi^3$. This term can facilitate a three-wave mixing (3WM) process where a single pump photon
        at roughly twice the signal frequency gives its energy to a pair of signal and idler photons~\cite{fadavi2023three, perelshtein2022broadband, malnou2021three}. Having the pump tone far from the signal frequency is beneficial, as then the strong tone at the output of the TWPA can be removed with a simple filter and one avoids any compression effects in later amplification stages. 
        
        If the signal and idler frequencies are the same, the resulting amplification is said to be \textit{degenerate} and only one of the signal quadratures is amplified, but possibly without any added noise. In our system, we typically employ 3WM TWPAs in the non-degenerate regime such that we can frequency multiplex the readout. However, by changing the flux bias current it is possible to tune the degeneracy to a frequency of interest without changes to the hardware.
        
\subsection{QPU control electronics}
\label{control_electronics_subsection}

    The microwave drive pulses are generated by conventional AC-coupled microwave arbitrary waveform generators (AWGs) operating at the qubit frequency band. The tunable coupler flux pulses are generated by a DC-coupled baseband AWG.
    The readout probe signals are generated and acquired by a conventional quantum analyzer operating at the frequency band of the readout resonators. The readout and drive instruments also provide a combined functionality that enables fast feedback, i.e., driving signals dependent on the readout result at time scales shorter than the qubit coherence times.
    The qubit flux and TWPA bias currents are generated by a DC voltage source which is connected to the QPU and TWPA devices via a cascade of low pass filters that include inline resistance to convert voltage to a stable direct current. 
    
    The electronics racks of the system include all the required auxiliary electronics for operating a full-stack quantum computer: uninterruptible power supply (UPS) to provide regulation and filtering of the mains power to the measurement electronics rack, a network remote configurable mains power distribution unit (ePDU), power supplies for the various readout amplifiers, a reference clock, power supplies for the DC sources, a main Linux host for running the control software, a Windows host for running the Bluefors software that controls the dilution refrigerator, several smaller specialised Linux hosts for instrumentation interfaces, a dedicated firewall, and a network switch that provides Ethernet connectivity to the hosting facility.

\section{Software}
\label{sec:software}

\begin{figure}[tb]
\centering
\includegraphics[width=0.8\textwidth]{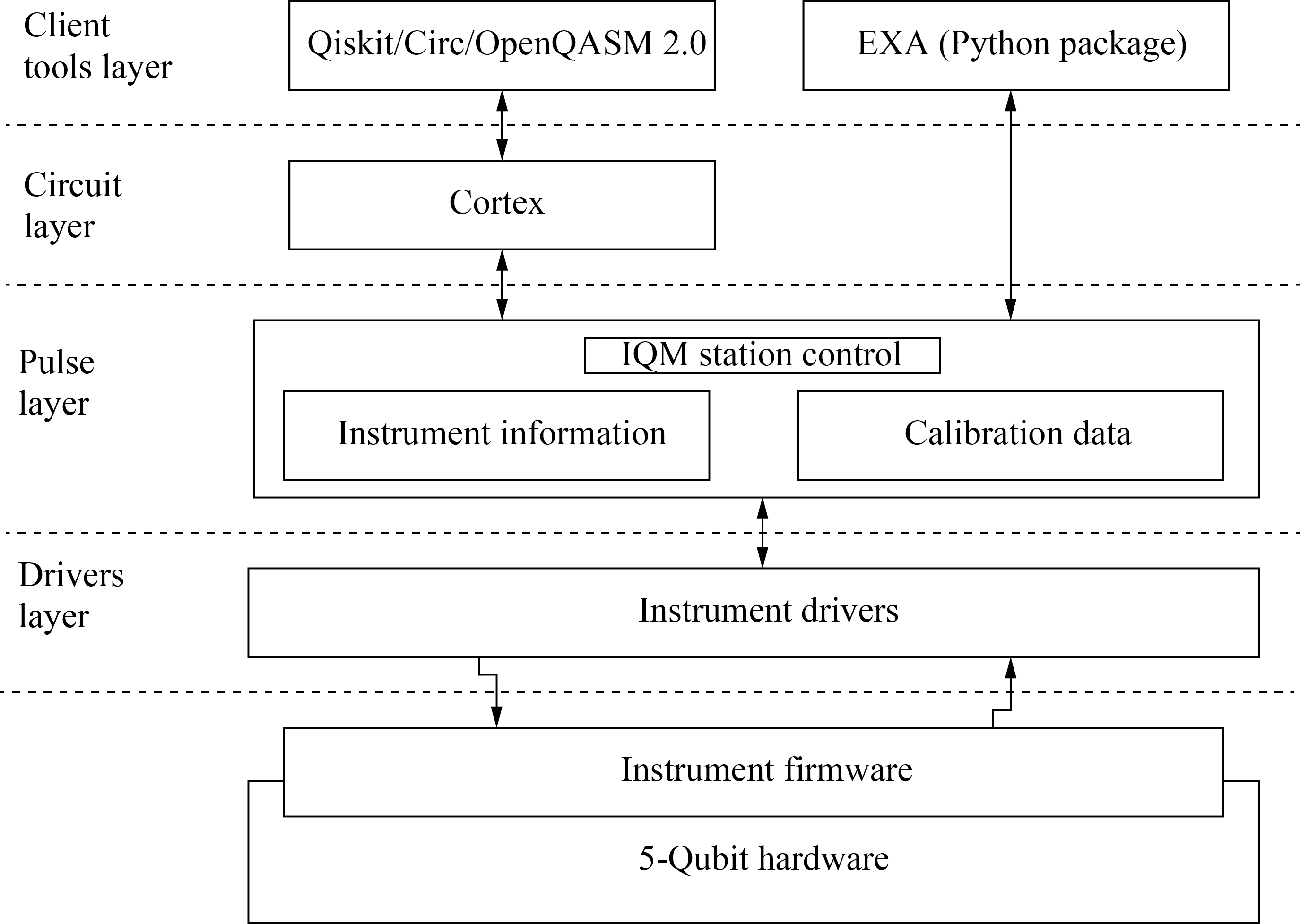}
\caption{The software layers and modules of our quantum computer control software stack.\label{fig:swstack}}
\end{figure}

The software stack of our quantum computer is divided into different functional layers presented in Fig.~\ref{fig:swstack}.
It can be interfaced in several ways based on the required level of access. The modules and interactions of the software stack are described in more details in the following subsections.

\subsection{Cortex}
Cortex is a set of software components for running quantum algorithms on our quantum computer. 
It is the highest level of abstraction in the control software stack of our on-premises quantum computers.
Cortex focuses on enabling computation for the end user, rather than experimenting with the behaviour of the individual elements of the quantum computer.

Cortex allows users to define and execute quantum algorithms on the quantum computer, expressed as quantum circuits using high-level frameworks and description languages such as Cirq, Qiskit, and OpenQASM 2.0. The input to the server is a computation job containing one or more quantum circuits to be executed, the number of shots, and possibly some other parameters. The job is queued for execution on the quantum computer. 
The results of the measurements of the circuits are returned when the job is completed.

\subsection{EXA}
EXA is a Python-based framework for characterising, calibrating, and controlling our quantum computer. EXA supports the execution of pre-defined experiments as well as the definition and execution of new custom experiments. An experiment unit combines different functionalities such as execution flow, data manipulation, analysis, and presentation, and it can also be built in a modular way using other experiments.

Using the EXA experiment library, users can create Jupyter notebooks or standalone Python applications e.g. to implement macro-like capabilities to simplify the control and measurement processes, eliminate standard repetitive operations using automated procedures, and develop entirely new experiments.

\subsection{IQM Station Control}
IQM Station Control takes care of low-level functionality such as housing instrument parameters and hardware drivers. It hides low-level hardware details from the higher-level components, EXA and Cortex.
Both Cortex and EXA communicate with Station Control service via its non-RESTful JSON (JavaScript Object Notation) HTTP (Hypertext Transfer Protocol) interface. The interface provides endpoints for performing various parameter sweeps and executing pulse schedules. In normal use, the user does not need to interact directly with the service.

Station Control uses device-specific drivers to further encapsulate the details of each instrument, including its low-level communication protocol.

\section{Applications to education}
\label{sec:education}

A small scale on-premises quantum computer is exceptionally useful for educational purposes. It facilitates hands-on experimentation, allowing students not only to run quantum circuits, but also to conduct pulse-level experiments, change the calibration, or connect external periphery. In general, applications to education can be sorted into two categories, (1) experiments/lab sessions that involve accessing the hardware physically or through the pulse-level interface, and (2) accessing the quantum computer through the circuit-level interface.

\subsection{Utilizing hardware access/pulse-level access for education}

Lab sessions that involve accessing the hardware physically or utilizing the pulse-level interface and changing the configuration or calibration of the device help engage students and provide additional learning opportunities that are pivotal in cultivating the next generation of quantum scientists and engineers. Below we will illustrate four examples of how an on-premises quantum computer can be used for this purpose in education.

\subsubsection{Exploring a quantum computer}

Access to a physical quantum computer enables the investigation of the setup of these machines. 
Together with the appropriate exercises and learning materials this kind of physical access bridges the gap between abstract quantum circuit descriptions and the actual superconducting quantum computer that executes them. Students experience first hand how the qubits are protected from the environment via cooling and magnetic shields, and also how information is exchanged with the classical world.

\subsubsection{Calibrating a quantum computer}
\begin{figure}
\centering
   \includegraphics[width=\linewidth]{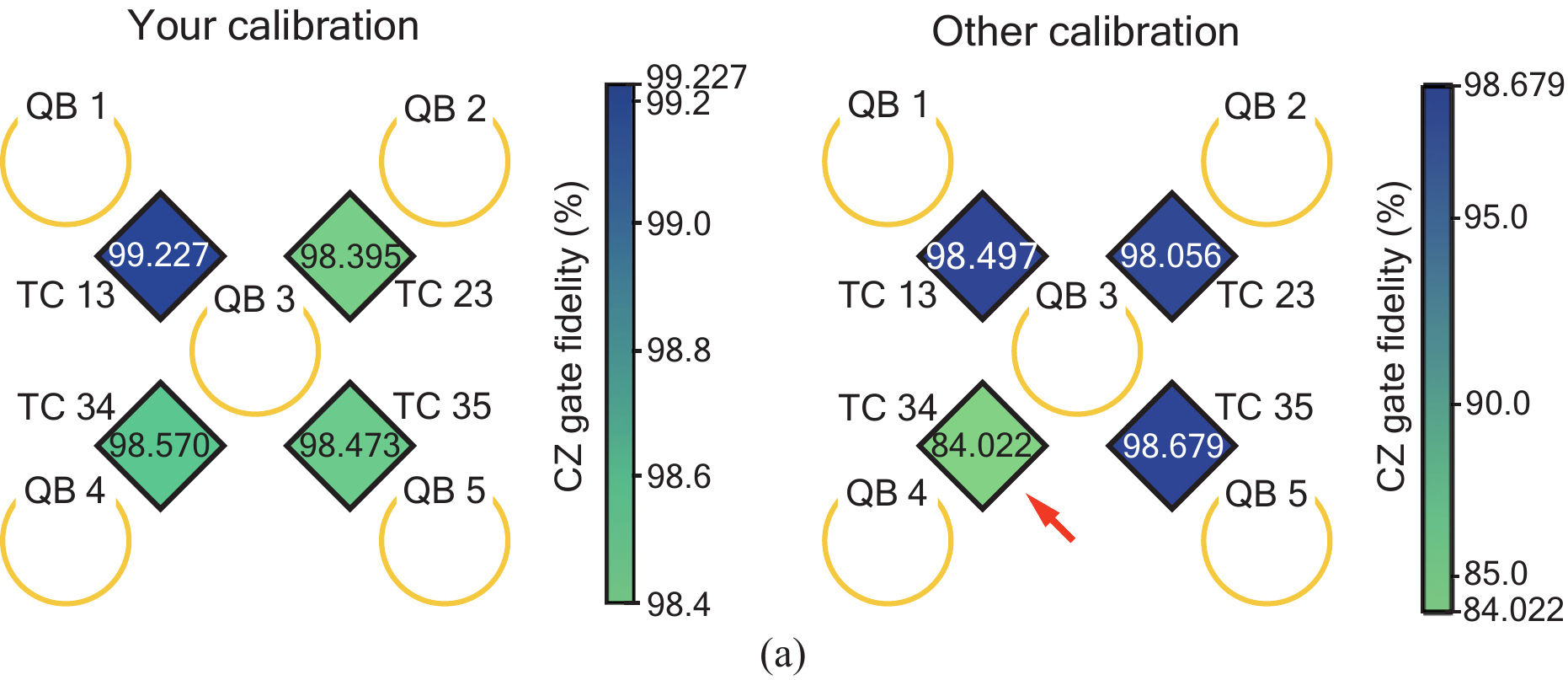}
   \includegraphics[width=\linewidth]{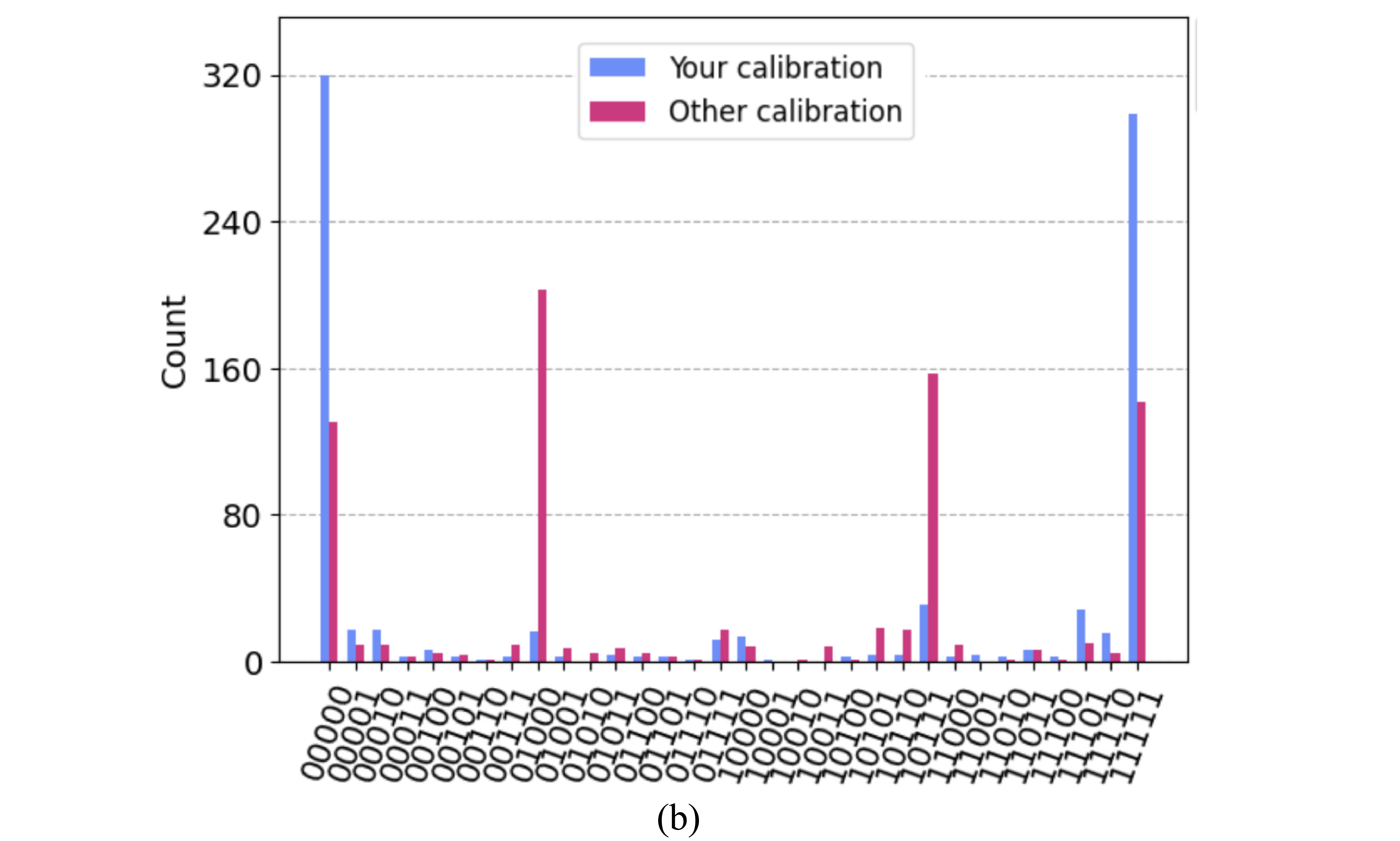}
   \caption{(a) Learners may compare their calibration with another. They notice the badly calibrated $CZ$ gate in the second calibration set, indicated by the red arrow, and investigate the cause. (b) Comparison of the 5-qubit GHZ state preparation fidelity using different calibration sets shown in (a).   \label{fig:calibration}}
\end{figure}

By utilizing the pulse-level interface, learners can explore the importance of calibration and create their own calibration sets. Calibration is required to find the optimal parameters to operate the QPU and is crucial for high fidelity operations. In a lab setting, the learners create and apply their calibration sets and compare outcomes to understand the impact of calibration on the results as shown in Fig.~\ref{fig:calibration}~(a). Figure~\ref{fig:calibration}~(b) shows the measurement outcomes of the 5-qubit GHZ state
\begin{equation}
    \ket{\Psi}=\frac{1}{\sqrt{2}}(\ket{00000}+\ket{11111})
\end{equation}
with two different calibration sets. 

Furthermore, learners can investigate the fidelity of  gates and the potential causes of discrepancies in calibration outcomes, encouraging students to critically analyze their results against provided benchmarks. Benchmarking can also expand to aspects such as $T_1$ and $T_2$ times or different strategies for assessing performance.

\subsubsection{Exploring control waveforms}
\begin{figure}
\centering
   \includegraphics[width=1\linewidth]{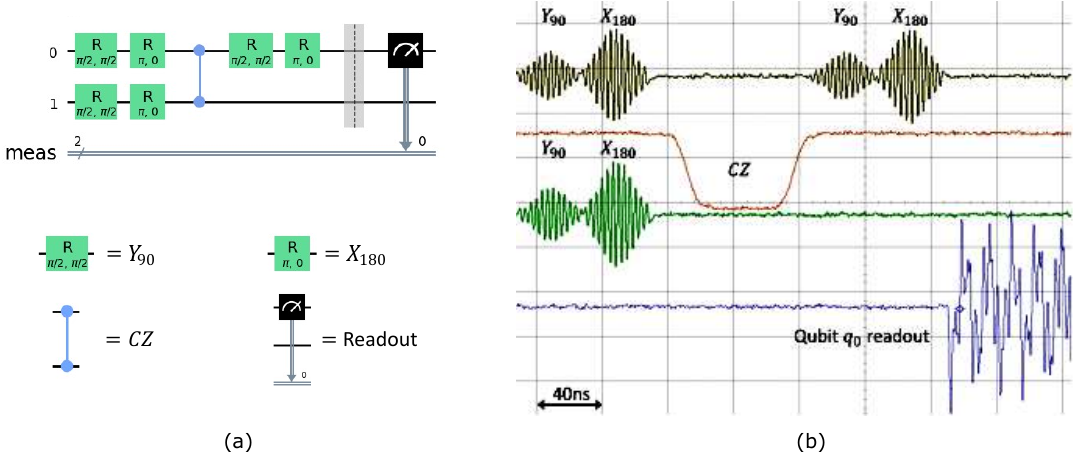} 
   \caption{(a) Transpiled circuit that creates a Bell state using native operations of our quantum computer. Note that this circuit has been chosen for illustrative purposes and has not been fully optimized for the given architecture. (b) Resulting pulse shapes as displayed on the oscilloscope interface. Horizontal and vertical axis depict time and voltage, respectively. Curves are shifted vertically for better visibility.\label{fig:oscilloscope}}
\end{figure}

With physical access to the device, students can also plug in selected peripheral devices such as oscilloscopes to further investigate the connection of software-defined operations and the physical implementation of quantum operations. Multiple smaller experiments guide the learners through investigating control pulse characteristics and qubit manipulation. For example, they execute multiple instructions that rotate the qubit state by different angles and measure the corresponding pulse shapes. As a more complicated example, learners can explore the control pulse schedules that result from multi-qubit circuits such as one that generates a Bell state $\ket{\Phi_+}\,{=}\,(\ket{00}+\ket{11})/\sqrt{2}$, as shown in Fig.~\ref{fig:oscilloscope}~(a). The resulting control and readout pulses measured with an oscilloscope are depicted in Fig.~\ref{fig:oscilloscope}~(b). Through this hands-on approach, learners will gain insight into the control electronics that enable superconducting quantum computing and the concepts of pulse control.

\subsubsection{Multi-level quantum hardware}

Direct hardware access allows one to investigate physical quantum systems beyond the two-level approximation defining a qubit. Here we demonstrate the state preparation and readout of the second excited state of a transmon. This example provides interested learners with a better understanding of the actual superconducting quantum hardware. Furthermore, it connects the educational value of the system with recent scientific results enabled by utilizing the multi-level nature of the transmon~\cite{PhysRevA.76.042319}, such as three-level (qutrit) quantum processors~\cite{PhysRevX.11.021010}, tunable coupler architectures using the second excited state~\cite{fabian_coupler} and fast single-qubit gates by the shortcuts-to-adiabaticity version of the stimulated Raman processes (STIRAP)~\cite{doi:10.1126/sciadv.aau5999}. 

\begin{figure}
\centering
   \includegraphics[width=.8\textwidth]{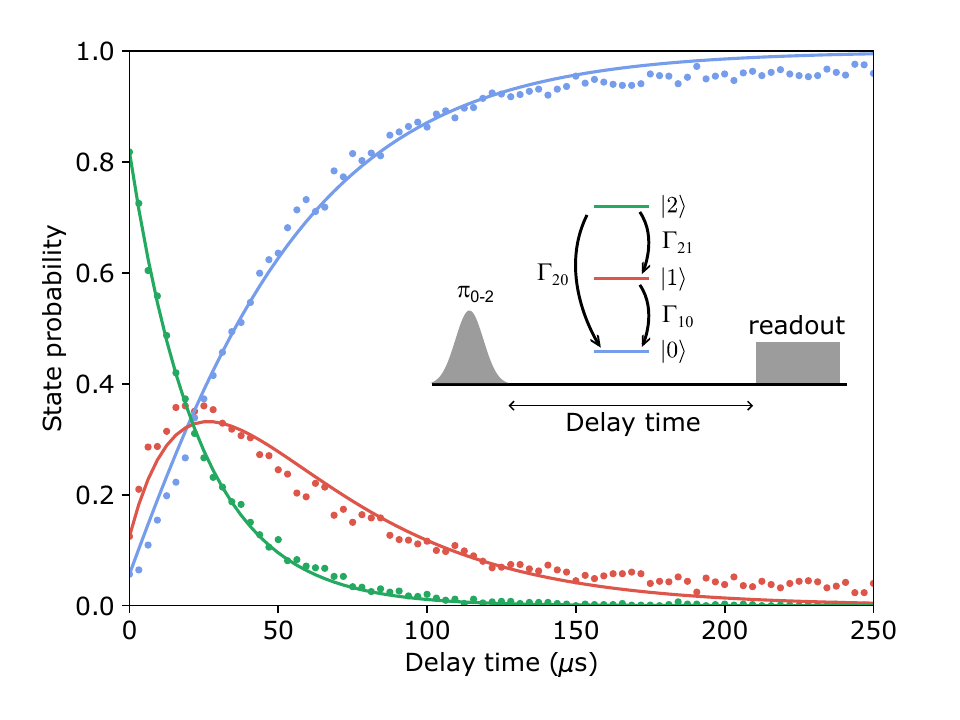}
   \caption{State preparation and relaxation dynamics of a qutrit. The inset displays the pulse sequence, which starts with preparing the qutrit in the second excited state, and the readout is performed after a variable delay time. The result of the single shot analysis is shown as dots, and the best fit to the Markovian model in the inset is displayed as solid curves with the relaxation rates $\Gamma_{21}$, $\Gamma_{20}$ and $\Gamma_{10}$ given in the text.\label{fig:qutrit}}
\end{figure}

In Fig.~\ref{fig:qutrit}, we display the results of an experiment addressing the relaxation dynamics of a transmon prepared in the $\ket{2}$ state by a calibrated $\pi_{0-2}$ pulse. The population then freely evolves and state discrimination is performed after a delay time (see inset). The observed evolution (dots) of the three states $\ket{0}$, $\ket{1}$ and $\ket{2}$ is then fitted (solid lines) to extract the relaxation timescales. We find $\Gamma^{-1}_{10}=44.4\pm1.3\,\mu$s, $\Gamma^{-1}_{21}=35.0\pm1.1\,\mu$s and $\Gamma^{-1}_{20}=69.2\pm4.1\,\mu$s, which together capture the expected dynamics of the qutrit, where the inequalities $\Gamma^{-1}_{10}>\Gamma^{-1}_{21}$ and $\Gamma^{-1}_{10},\Gamma^{-1}_{21}<\Gamma^{-1}_{20}$ demonstrate the role of the transition matrix elements when describing the relaxation of a quantum system~\cite{PhysRevLett.114.010501}. We conclude that pulse-level access enables the direct investigation and control of the superconducting quantum hardware beyond what higher abstraction layers can provide, making it a vital tool for educational programs targeting quantum hardware.

\subsection{Utilizing circuit-level access for education}

A lot of interest goes into investigating, creating and improving NISQ algorithms. This subsection will demonstrate the use of gate-based access in educational settings by providing different examples that can be practiced by learners. The output of each quantum circuit is illustrated by executing it with our superconducting quantum computer. 

Because of the unavoidable presence of errors during algorithm execution on NISQ devices such as our quantum computer, the results are expected to differ from the ideal, noiseless ones. Importantly, there are strategies, often collectively referred to as ``quantum error mitigation'', that can be employed, either individually or in combination with each other, to reduce the effects of errors on the execution of a desired algorithm~\cite{ref:cai2022}. In particular, in the examples that will follow, we make use of techniques belonging to three different classes.
\begin{itemize}
    \item ``Error suppression'' techniques aim to modify and reduce the effects of errors at the level of each single circuit run. A prominent example is the randomized compiling (RC) technique~\cite{hashim2021, wallman2016} that, through random Pauli twirling, effectively converts problematic coherent errors (e.g. systematic overrotations associated with a given gate) to stochastic errors, which add up more favourably and whose effects are easier to mitigate further. 
    
\item ``Readout error mitigation'' (REM) techniques target errors that occur during the measurement of the qubits~\cite{ref:nation2021, ref:cai2022}. In general, they consists of two steps. At first, a set of simple and shallow characterization circuits is executed on the hardware, in order to determine the properties and the magnitude of the readout errors. As a simple example, one might want to measure what is the probability that a given qubit prepared, say, in the $|0\rangle$ state is incorrectly measured to be in $|1\rangle$. Once this information is gathered, the desired algorithm is executed and its raw results are post-processed in order to compensate for the readout errors. In this paper, we have mitigated readout errors using correlated readout error mitigation calibrated with 10,000 shots per basis state.

\item A third class of ``error mitigation'' techniques mainly target errors happening at the gate level, i.e. during the execution of the bulk of the circuits. This is generally achieved by executing different variants of the desired quantum circuit and by combining their output via classical post-processing, leading to a (potentially significant) run time overhead~\cite{ref:cai2022}. Despite this trade-off between quality and speed, the implementation of error mitigation strategies is a key element for successful algorithm execution in the NISQ era. One simple-yet-effective technique to mitigate gate errors is known as zero noise extrapolation (ZNE)~\cite{ref:temme2017}, which is based on the idea of artificially increasing the noise level and then making use of noisier results to extrapolate back to the noiseless limit. Open source and educational implementations of several other techniques can be found, for example, in~\cite{ref:larose2022}. 
\end{itemize}

\subsubsection{Violation of the CHSH inequality}

The Nobel Prize in Physics 2022 was awarded jointly to Alain Aspect, John F. Clauser and Anton Zeilinger ``for experiments with entangled photons, establishing the violation of Bell inequalities and pioneering quantum information science''~\cite{bib:nobel2022}. Bell's theorem claims that correlations of measurement outcomes of two experimenters separated from one another have an upper bound if nature follows the principle of local realism~\cite{ref:bell1964}. Suppose two parties, often called Alice and Bob, are located far away from each other in the experiment. They both have two observables they can measure of some signal that comes to them, but they cannot measure both simultaneously. Bell-type inequalities define correlators over the possible measurement settings and their outcomes, assuming that the measurements happen so fast that no communication is possible between the measurement devices of Alice and Bob due to the finite speed of light~\cite{scarani2019bell}. A Bell inequality then separates so-called local probability distributions from non-local distributions, that have stronger correlations than the local distributions. The most famous Bell-type inequality is the CHSH inequality~\cite{ref:chsh1969}, which is straightforward to test with two qubits on a quantum computer.  The measurements of the qubits are definitely not space-like separated events, though. The inequality is given by the formula:
\begin{equation}
    -2 \leq E(QS)+E(RS)+E(RT)-E(QT) \leq 2,
    \label{eq:chshclassical}
\end{equation}
if the local realism is true. Here the observables are
$$
Q=X_1 
\quad R=Z_1
\quad S=Z_2
\quad T=Z_2
$$
where index $1$ refers to Alice's and $2$ to Bob's qubit and $E(A)$ stands for the expectation value of~$A$.

Let us now consider how these correlations behave in quantum mechanics, and evaluate the above expectation values in the state
\begin{equation}
\ket{\Psi(\theta)} = R_y(\theta)_1 (\ket{00}+\ket{11})/\sqrt{2},
\end{equation}
where $R_y(\theta)= \exp(-i \theta Y/2)$. Calculating the expectation values as described at the end of section~\ref{sec:introduction}, we obtain
\begin{equation}
    E(QS)=E(RT)= \cos \theta,\quad  E(QT)=-E(RS) = \sin \theta,
\end{equation}
and thus
\begin{equation}
    E(QS)+E(RS)+E(RT)-E(QT) 
    = 2\sqrt{2} \cos(\theta+\pi/4).
    \label{eq:chshcomb}
\end{equation}
Clearly the CHSH inequality in Eq.~\eqref{eq:chshclassical} is violated in the state $\ket{\Psi(\theta)}$ iff
$$
\theta \in \left(\frac{\pi}{2}, \pi \right) \sqcup \left(\frac{3\pi}{2},2\pi\right),
$$
which demonstrates that quantum mechanics is not compatible with local realism.

\begin{figure}[!ht]

    \subfloat[\label{subfig:bell_pair_circuit}]{%
    \hspace{2.9cm}  \includegraphics[width=0.32\textwidth]{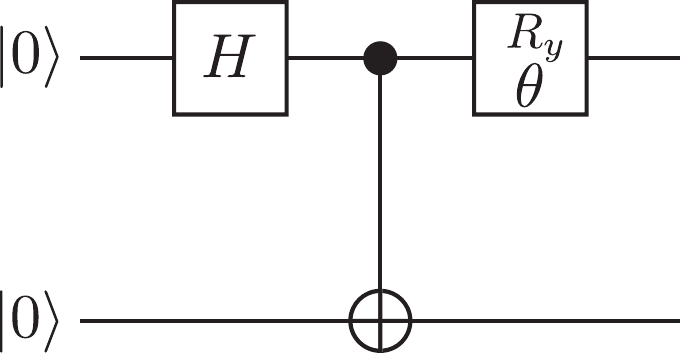}
    }
    \\
    \subfloat[\label{subfig:CHSH_scan}]{%
      \includegraphics[width=0.8\textwidth]{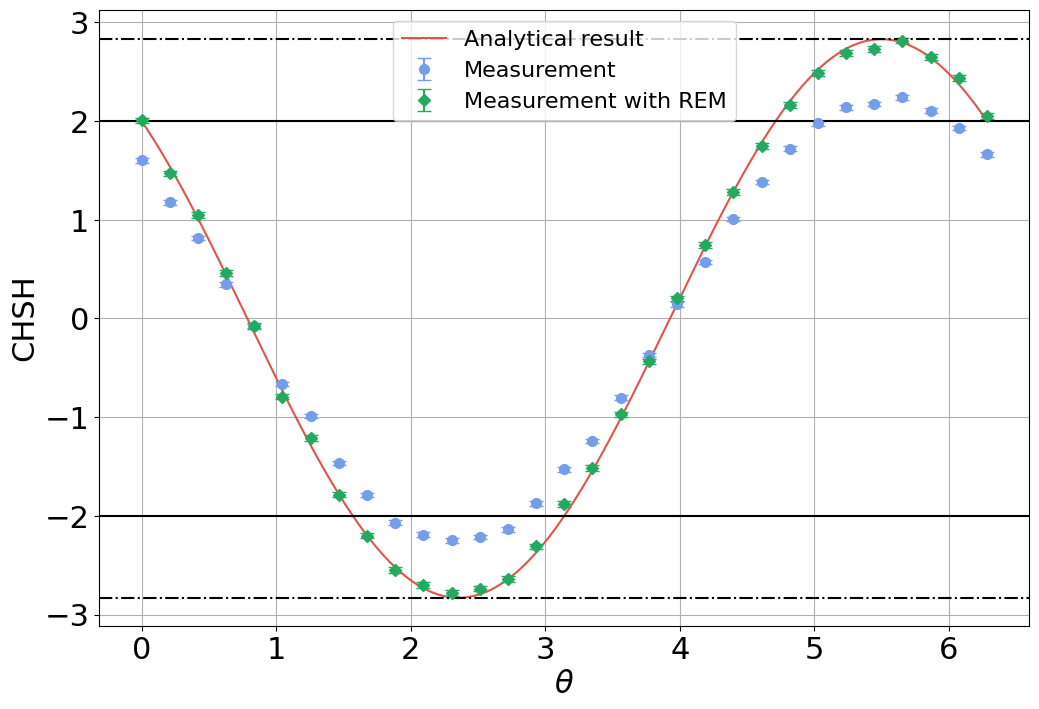}
    }
    \caption{(a) Quantum circuit that prepares $\ket{\Psi(\theta)}$. (b) Expectation value of the CHSH observable Eq.~\eqref{eq:chshcomb} in the state $\ket{\Psi(\theta)}$, plotted over the $y$ rotation angle $\theta$. The raw experimental data points are shown with blue dots; the implementation of readout error mitigation (green crosses) brings them closer to the ideal noiseless result (red curve). There are two regions outside the black horizontal lines, where the CHSH inequality is violated and hence non-locality is demonstrated. The statistical uncertainty is so small that the error bars remain within the markers. 
    }
    \label{fig:chsh}
  \end{figure}

To experimentally test this prediction, we will execute a parameterised quantum circuit given in Fig.~\ref{fig:chsh}~(a) to create the state $\ket{\Psi(\theta)}$ and measure the relevant expectation values for multiple values of~$\theta$.
Figure~\ref{fig:chsh}~(b) shows how the CHSH observable oscillates as we rotate the state, and violates the equality as predicted. The statistical uncertainty due to the finite amount of shots is shown by the error bars, which correspond to one standard deviation. They have been determined by means of bootstrapping, i.e. by classically resampling several times the probability distribution obtained from each circuit run, thus obtaining a set of reprocessed results that can be used to estimate confidence intervals. We use this method throughout the paper.

\subsubsection{5-qubit GHZ state, decoherence and Mermin's inequality}

\noindent
{\bf 5-qubit GHZ state}

\begin{figure}
\centering
\includegraphics[width=12cm]{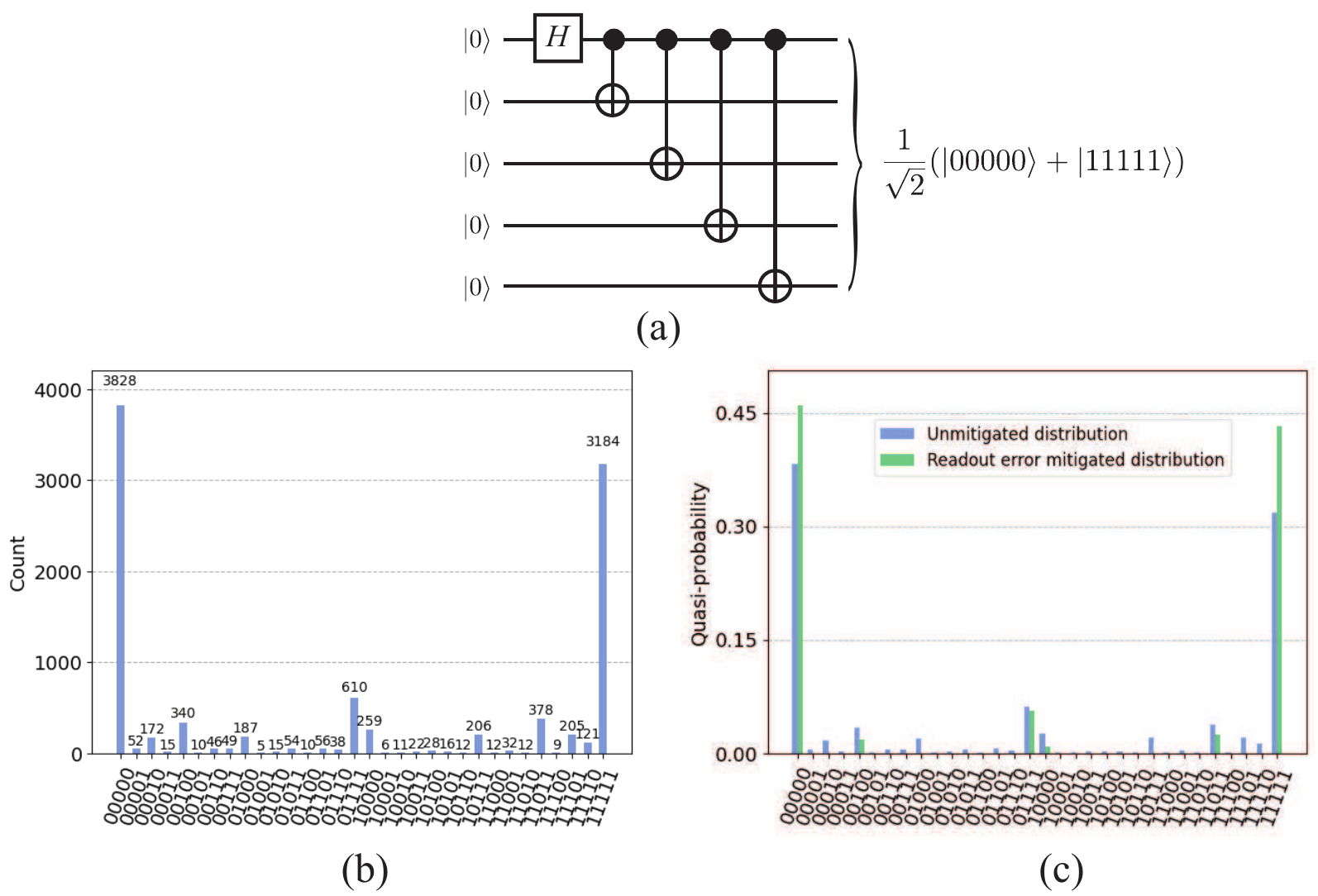}
\caption{(a) Quantum circuit to implement the 5-qubit GHZ state. (b) Output of 5-qubit quantum computer. (c) Output after readout error mitigation is applied. 
\label{fig:ghz5}
}
\end{figure}

Let us prepare a maximally entangled 5-qubit state and see what we can do with it. The 5-qubit GHZ state \cite{ref:GHZ1989}
\begin{equation}
\ket{\Psi_5} = \frac{1}{\sqrt{2}}(\ket{00000}+\ket{11111})
\end{equation}
is one of the maximally entangled 5-qubit states. This state is obtained by applying the quantum circuit in Fig.~\ref{fig:ghz5}~(a) on $\ket{00000}$. Figure~\ref{fig:ghz5}~(b) shows the output histogram of our quantum computer obtained with 5,000 shots, while Fig.~\ref{fig:ghz5}~(c) shows the output after readout error mitigation is applied.  
\vspace*{.3cm}

\noindent
{\bf Entanglement, mixed state and decoherence}

\begin{figure} 
\centering
\subfloat[]{
\includegraphics[width=0.95\columnwidth]{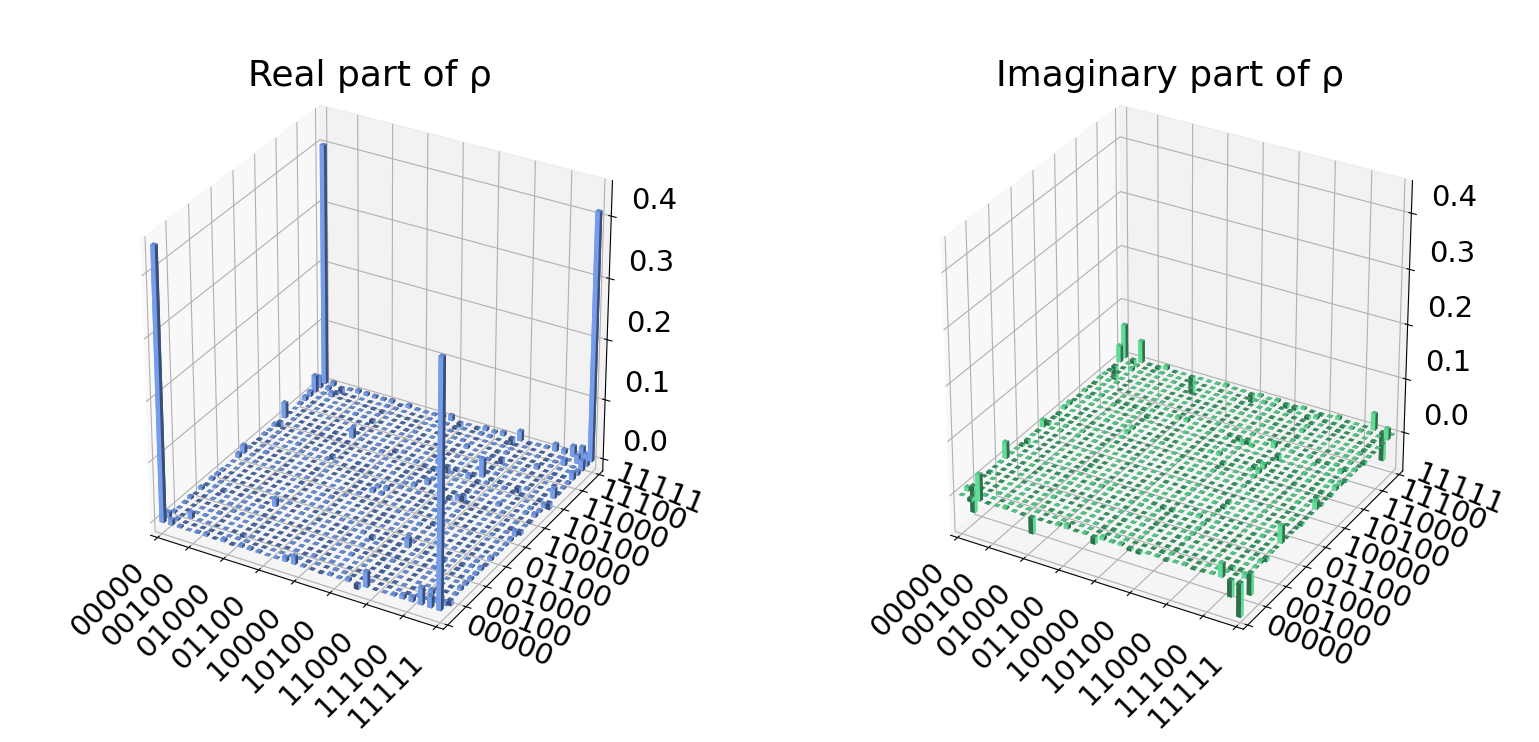}}
\\
\subfloat[]{\includegraphics[width=0.95\columnwidth]{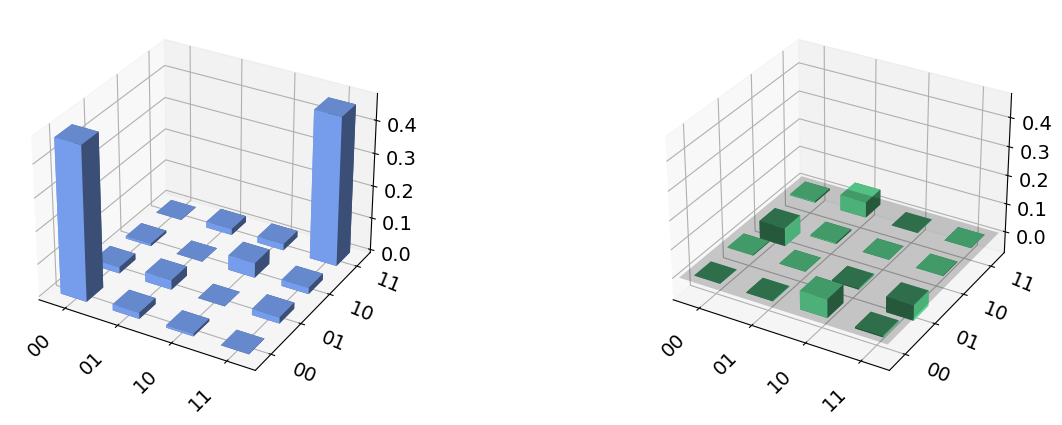}}
\\
\subfloat[]{\includegraphics[width=0.95\columnwidth]{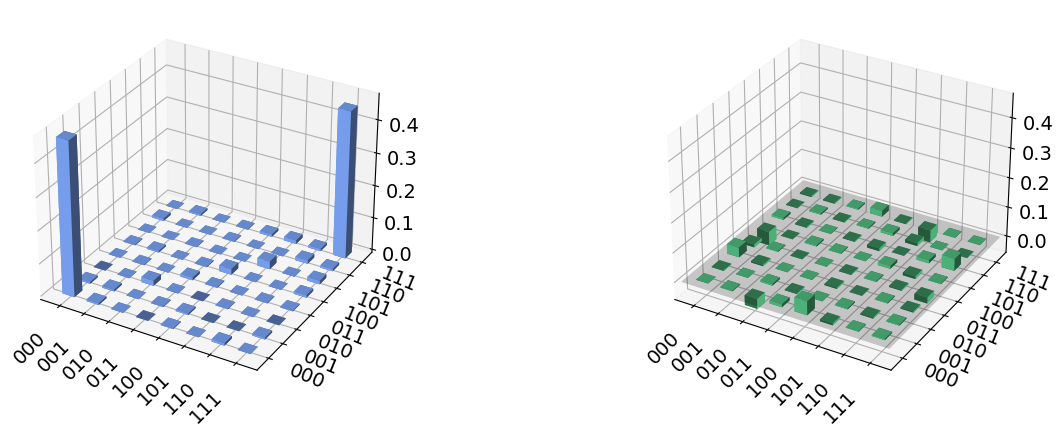}}
\caption{State tomography of (a) 5-qubit GHZ state, (b) qubits 12 (principal system) and (c) qubits 345 (environment). Theoretically (a) is a pure state with rank 1 while (b) and (c) are mixed states with rank 2. We used local readout error mitigation and 3500 shots per measurement basis for the tomography.}
\label{fig:GHZtomography}
\end{figure}

Let us separate the 5-qubit system into subsystems made of qubits 12 and qubits 345. Suppose one measures an observable $O$ associated with subsystem 12. The expectation value of $O$ with respect to $\ket{\Psi}$ is
\begin{equation}
    \langle O \rangle = \bra{\Psi}(O \otimes I_8)\ket{\Psi}= \frac{1}{2} \bra{00}O\ket{00}+    \frac{1}{2} \bra{11}O\ket{11},
\end{equation}
which is an expectation value with respect to a mixed state, even though the total system is in a pure state. This is directly demonstrated by evaluating the density matrices of (a) the 5-qubit GHZ state, (b) the subsystem 12 and (c) the subsystem 345 as
\begin{equation}
\rho_\mathrm{GHZ} = \frac{1}{2}
\begin{pmatrix}
1&0&\ldots &0&1\\
0&0&\ldots &0&0\\
\vdots& &\vdots& &\vdots\\
0&0&\ldots &0&0\\
1&0&\ldots &0&1
\end{pmatrix},
\end{equation}
\begin{eqnarray}\label{eq:sub12}
\rho_\mathrm{12} &=& \sum_{i,j,k \in\{0,1\}}
(I_2 \otimes I_2 \otimes \bra{ijk})\rho_\mathrm{GHZ}
(I_2 \otimes I_2 \otimes \ket{ijk})=\frac{1}{2} \mathrm{diag}(1,0,0,1)\\
\rho_\mathrm{345} &=& 
 \sum_{i,j \in\{0,1\}}
(\bra{ij}\otimes I_2 \otimes I_2 \otimes I_2)\rho_\mathrm{GHZ}
(\ket{ij} \otimes I_2 \otimes I_2 \otimes I_2)\nonumber\\
&=&\frac{1}{2}\mathrm{diag}(1,0,0,0,0,0,0,1),
\end{eqnarray}
respectively. Figure~\ref{fig:GHZtomography} shows the result of quantum state tomography for (a) the GHZ state (b) subsystem 12 and (c) subsystem 345, which are obtained experimentally with the Qiskit state tomography algorithm. 
The subsystem 12 is in a mixed state since it cannot access the information of the subsystem 345 and vice versa. 

This situation models decoherence. It is often said that interaction of a quantum system with environment causes entanglement between the two systems, by which a initial pure state of the system becomes a mixed state. This explanation of decoherence is often difficult to understand for beginners. Let us call the subsystem 12 the principal system while subsystem 345 the environment. The total system, started in a pure tensor product state $\ket{00} \otimes \ket{000}$, evolves to the entangled GHZ state under the unitary time evolution given by Fig.~\ref{fig:ghz5}~(a). The total system is still in a pure state. Although the principal system was in a pure state $\ket{00}$ in the beginning, it is now entangled with the environment, and thus in a mixed state $\rho_{12}$ if the environment is ignored (i.e. traced out). In other words, the GHZ state is a purification of $\rho_{12}$.

 It is interesting to evaluate the von Neumann entropy $S(\rho) =-\mathrm{tr}\, \rho \log_2 \rho$ of these states. Table~\ref{table:vN} shows both the theoretical predictions and the experimental results along with the upper bound saturated by the uniformly mixed state. Observe that the entropy, which is called the entanglement entropy in this context, is theoretically the same for $\rho_{12}$ and $\rho_{345}$. Usually, entropy is proportional to the system size but entanglement entropies derived from a pure state $\rho_\mathrm{GHZ}$ are identical even though the subsystem sizes are different.
\begin{table}
\centering
\captionsetup{width=\textwidth}
\begin{tabular}{|c|c|c|c|}
\hline
 & Theory & Experiment& Upper bound\\
 \hline
GHZ &  0& 0.925  & 5\\
\hline
$\rho_{12}$ & 1 & 1.306& 2  \\
\hline
$\rho_{345}$ & 1 & 1.386& 3 \\
\hline
\end{tabular}
\caption{Theoretical and experimental values of the von Neumann entropy of the 5-qubit GHZ state, the subsystem 12 and the subsystem 345. Experimental data in Fig.~\ref{fig:GHZtomography} has been employed. The right column shows the upper bound of entropy, which is saturated by the uniformly mixed state.\label{table:vN}}
\end{table}

\vspace*{0.3cm}
\noindent{\bf Violation of Mermin's inequality}

Let us show next that Mermin's inequality is violated by the 5-qubit GHZ state. Mermin's inequality is regarded as a generalization of the CHSH inequality to multi-qubit systems \cite{mermin}. The Mermin polynomial for a 5-qubit system is defined as~\cite{mermin5}
\begin{eqnarray}
    M_5&=& X_1X_2X_3X_4X_5\nonumber\\
    & & -(Y_1Y_2X_3X_4X_5 + 9~\mbox{permutations})\nonumber\\
    & & +(Y_1Y_2Y_3 Y_4X_5+4~\mbox{permutations}).
\end{eqnarray}
It is known that Mermin's inequality $E(M_5) \leq 4$ is satisfied if the local realism holds. On the other hand, quantum theory predicts $E(M_5) \leq 4^2=16$, where the upper bound is saturated if the state is maximally entangled.

Let us evaluate $E(M_5)$ in the 5-qubit GHZ state. Since the GHZ state is symmetric with respect to permutations of qubits we only need to evaluate three monomials:
\begin{eqnarray}
&\bra{\Psi}
X_1 X_2 X_3 X_4 X_5
\ket{\Psi}=1,&\nonumber\\
&\bra{\Psi}
Y_1 Y_2 X_3 X_4 X_5
\ket{\Psi}=-1,&\\
&\bra{\Psi}
Y_1 Y_2 Y_3 Y_4 X_5
\ket{\Psi}=1,&\nonumber
\end{eqnarray}
theoretically. Then we find
\begin{eqnarray}
   \bra{\Psi} M_5 \ket{\Psi} = 1- 10 \times (-1) + 5 \times 1=16,
\end{eqnarray}
and thus the GHZ state $\ket{\Psi}$ saturates the upper bound. Let us now confirm this prediction with our 5-qubit quantum computer.


The monomials of the Mermin polynomial can be measured as expectation values as described in section~\ref{sec:introduction}. Since the qubits and gate fidelities of a NISQ quantum computer are not homogeneous, we must measure all 16 monomials separately. The estimation of $\bra{\Psi} M_5 \ket{\Psi}$ obtained with our quantum computer is presented in Table~\ref{table:mermin}, and clearly rules out local realism, in favour of quantum theory. 

\begin{table}
\centering
\captionsetup{width=\textwidth}
\begin{tabular}{|c|c|c|c|}
\hline
Observable & Estimate without REM & Estimate with REM & Theoretical\\
\hline
$X_1 X_2 X_3 X_4 X_5$ & $0.5538$  & $0.9606$ & $1$\\
\hline
$Y_1 Y_2 X_3 X_4 X_5$ & $-0.4955$ & $-0.8624$ & $-1$\\
\hline
$Y_1 Y_2 Y_3 Y_4 X_5$ & $0.4398$ &  $0.7689$ & $1$\\
\hline
$M_5$ & $7.708$ & $13.43$ & $16$\\
\hline
\end{tabular}
\caption{
Estimated value of the Mermin polynomial in the prepared 5-qubit GHZ state.
The 16 terms of the polynomial were estimated using 10,000 shots each. The three topmost rows in the table represent the average of all the permutations of the Pauli operators in the given observable.\label{table:mermin}}
\end{table}

\subsection{Maxcut Problem}
\label{sec:va}

Many well known quantum algorithms, such as Shor's and Grover's algorithms, require a large number of qubits and fault-tolerant error correction for useful quantum computation. This places their practical execution beyond the capabilities of currently available NISQ computers. In contrast, variational quantum algorithms are more suited for the NISQ computer at our hand, in that they can be executed with a currently available number of qubits without quantum error correction. 

Variational quantum algorithms involve an optimization process, where classical computer seeks for the optimal parameters of a quantum circuit so that the expectation value of a Hamiltonian, representing the cost function, evaluated with a quantum computer using the resulting state is minimized. The parameters in the circuit are iterated many times until the expectation value hits the minimum. Variational algorithms have many use cases e.g. in mathematics, chemistry, finance and industrial optimizations. We will introduce application of a variational algorithm, called QAOA (Quantum Approximate Optimization Algorithm), to a combinatorial problem called the Maxcut problem in this subsection. Another variational algorithm VQE (Variational Quantum Eigensolver) will be introduced in section 5.3.

Suppose there is a graph $G$ with $n$ nodes. There are edges between some pairs of nodes. In the Maxcut problem, one aims to partition the nodes of $G$ into a disjoint union $A \sqcup B$ such that the number of edges connecting nodes in $A$ and $B$ is maximized. 

To solve this problem, we introduce an Ising Hamiltonian
\begin{equation}
    H=\sum_{i<j} J_{ij} Z_iZ_j,
\end{equation}
where $i$ and $j$ denote the nodes. 
The coupling strength is $J_{ij}=1$ if there is an edge between nodes $i$ and $j$, while $J_{ij}=0$ if there is no edge. Suppose $\ket{0}$ is assigned to nodes in group $A$, while $\ket{1}$ to nodes in $B$. 
If nodes $i$ and $j$ belong to different groups, the edge $ij$ contributes $-1$ to the Hamiltonian, while if they belong to the same group, the edge contributes $+1$. There is no contribution if there is no edge connecting nodes $i$ and $j$. Thus maximizing the number of edges connecting nodes in different groups reduces to minimizing the expectation value of the Hamiltonian $H$ acting on an $n$-qubit system.

In fact, the above problem may be implemented with $(n-1)$-qubit system. Suppose we find a solution $A$ and $B$ of a Maxcut problem of a given graph. Then interchange of $A$ and $B$ is also a solution of the same problem.
Accordingly we are free to assign $\ket{1}$ to the $n$th qubit, for example, without loss of generality. With this choise, $J_{in}Z_i Z_n$ becomes $-J_{in}Z_i$. The modified Hamiltonian 
\begin{equation}
    H'= \sum_{1\leq i<j \leq n-1} J_{ij}Z_i Z_j - \sum_{i=1}
    ^{n-1} J_{in}Z_i
\end{equation}
is implemented with an $(n-1)$-qubit system.

\begin{figure}
    \centering
    \includegraphics[width=0.4\columnwidth]{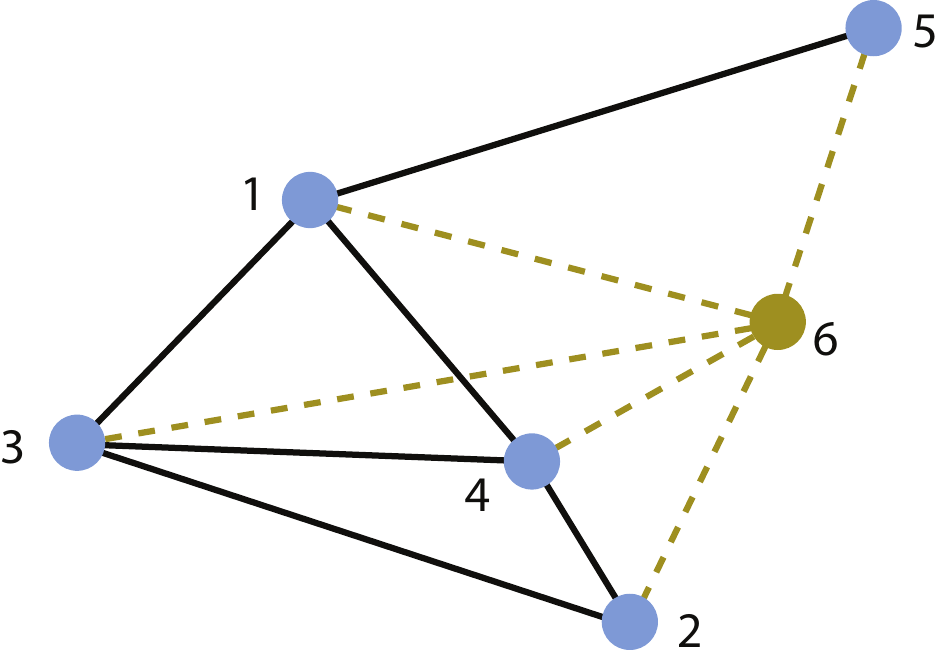}
    \caption{Maxcut problem for six nodes. Node 6 is a fictitious qubit in $\ket{1}$ state while rest are physical qubits. Solid lines represent physical couplings while dashed lines fictitious couplings.}
    \label{fig:Maxcut6}
\end{figure}

Let us consider a graph in Fig.~\ref{fig:Maxcut6} with 6 nodes for definiteness, where the node 6 is a virtual node in $\ket{1}$ state while the rest are physical. By fixing the state of node 6 to $\ket{1}$, the relevant Hilbert space is $\mathrm{Span}(\{\ket{i_1\, i_2\, i_3\, i_4\, i_5}\ket{1}\})$, which may be implemented with our 5-qubit quantum computer. 

\begin{figure}
\subfloat[]{
\includegraphics[width=0.65\columnwidth]{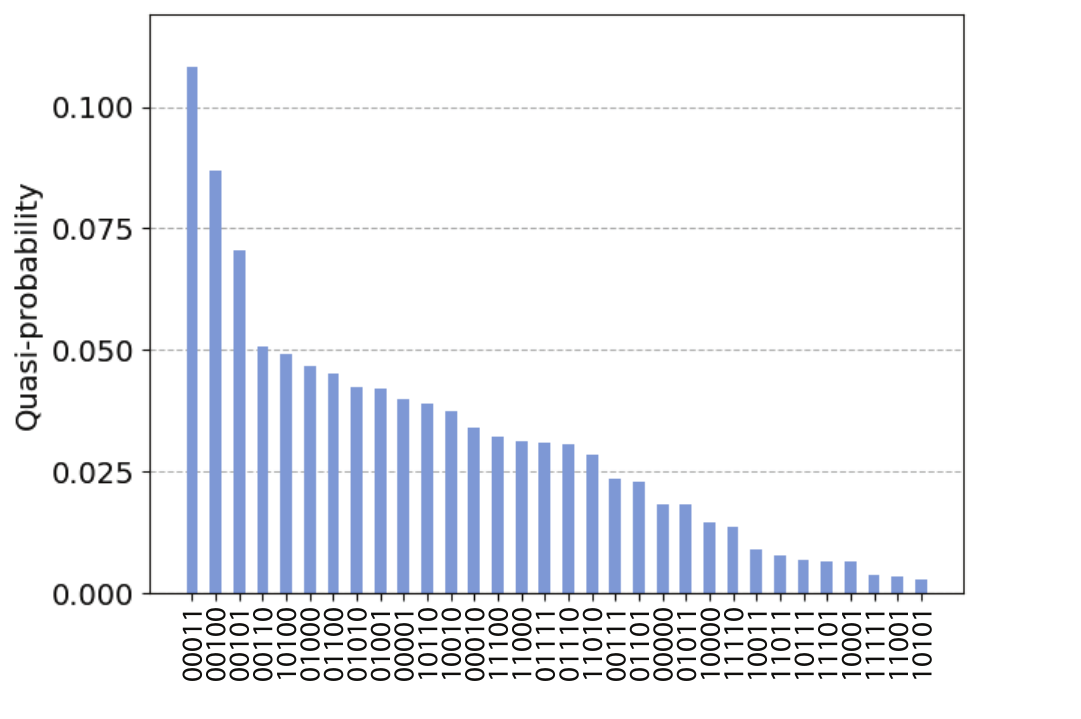}}
\subfloat[]{
\includegraphics[width=0.34\columnwidth]{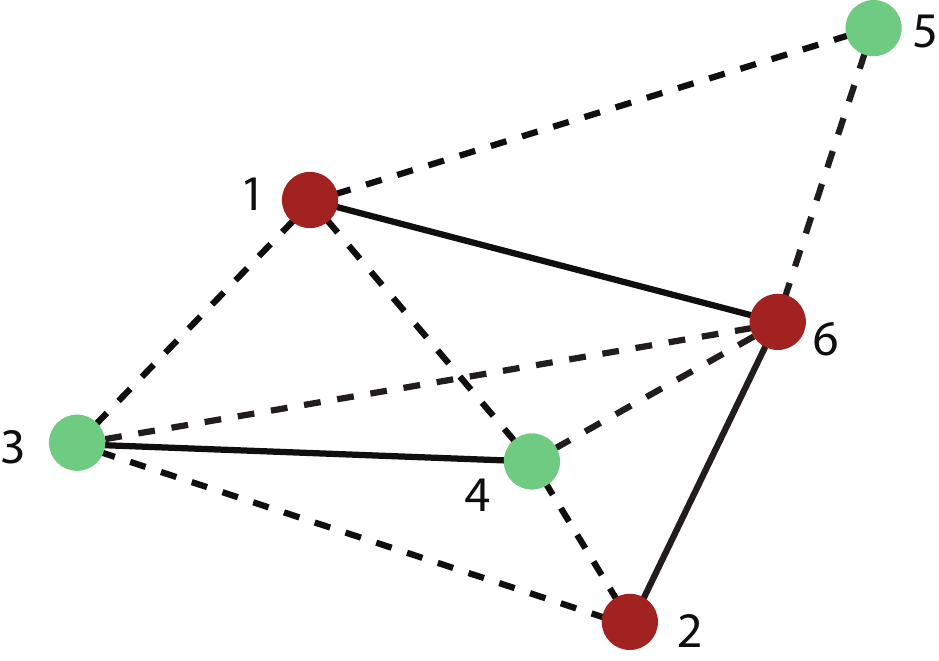}}
{\caption{(a) Measurement outcomes of our 5-qubit quantum computer executing QAOA for the Maxcut problem. (b) The solution of the Maxcut problem; we achieve eight cuts, which is the maximum for this graph. The solution $\ket{00011}$ has divided the nodes to groups 0 (green) and 1 (red), such that the number of edges connecting nodes from different groups is maximized. Solid lines show edges connecting nodes in the same group, while eight dashed lines are edges connecting nodes in different groups.}
\label{fig:Maxcut_output}}
\end{figure}

Figure~\ref{fig:Maxcut_output} shows the solution of the Maxcut problem experimentally obtained for the graph  Fig.~\ref{fig:Maxcut6}. The solution is read from the most probable state of the probability distribution. We used single-layered QAOA ansatz and $10,000$ measurements per optimization step and readout error mitigation. Using only a single layer of the QAOA ansatz means the algorithm has only a small number of gates and can be executed in a short time. This implies it is an approximate algorithm, and while it will output the correct solution, there will also be sizable probabilities of wrong states in the distribution. The algorithm can be more accurate and precise by increasing the layer depth, but then we start to accumulate more errors during the execution of the algorithm, which again introduces erroneous solutions to the output distribution of the quantum computer. The optimal depth depends in general on the problem instances and the error-levels of the quantum computer, and is a key implementation detail relevant for the performance of quantum algorithms on practical problems. 
\\

\noindent
{\bf Q-score}

Solving the Maxcut problem has been recently adopted as a benchmark for the practical capabilities of a quantum computer~\cite{martiel_2011}. The Q-score of a quantum computer equals the size of the graphs, whose Maxcut problem can be sufficiently solved. The obtained cost, i.e. the average number of cut edges, of the solution has to be above certain threshold. Specifically, one has to find the cost of a graph to be above 0.2 on a scale where 0 corresponds to random solution and 1 to ideal solution. The graphs chosen for the benchmark are random Erd\"os-R\'eny graphs with 50\% edge-probability between nodes.

\begin{figure}
    \centering
    \includegraphics[width=8cm]{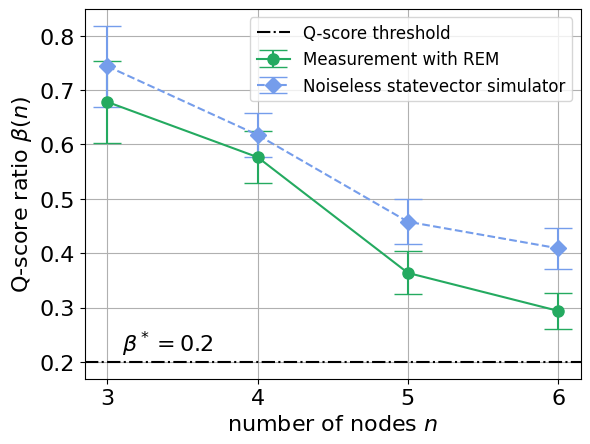}
    \caption{Q-score approximation ratios $\beta(n)$ for our 5-qubit quantum computer. Ratios above the threshold 0.2 pass the Q-score benchmark. Results of noiseless simulator are shown to highlight the approximate nature of QAOA; with one layer ansatz, the limited expressivity leads to results well below optimal ratio of $1.0$ even without noise. We employ the virtual node technique and readout error mitigation. The Q-score ratios are averages over 100 random Maxcut problems. We used 2048 shots per optimization step in QAOA.}
    \label{fig:Qscore}
\end{figure}

We present a comprehensive Q-score benchmark on our 5-qubit quantum computer. By employing the virtual node technique, we can solve the $n$-node graph with $n-1$ qubits. Figure~\ref{fig:Qscore} displays Q-score results up to five physical qubits.

\section{Applications to research}
\label{sec:research}

Small-scale quantum computers have been used in many different areas in scientific research including but not limited to physics, chemistry and mathematics. Current trend in NISQ computer research is, no doubt, toward scaling up physical qubits for commercial use of quantum computers. Nonetheless,~\cite{bib:survey} reports that there are many research papers published in scientific journals, which demonstrate ``proofs of principle'' of scientific ideas with a small-scale quantum computer.   

We illustrate three such examples from physics, mathematics and chemistry in this section, and demonstrate them with our 5-qubit superconducting quantum computer.

\subsection{Simulating Neutrino Oscillations} 

It is known that there are at least three types (i.e. flavors) of neutrinos in Nature. They are called $\nu_e, \nu_{\mu}$ and $\nu_\tau$. Masses of these neutrinos are not diagonal in the flavor basis $\{\ket{\nu_e}, \ket{\nu_\mu}, \ket{\nu_\tau}\}$. Let us call the eigenvectors of the mass matrix as $\{\ket{\nu_1}, \ket{\nu_2}, \ket{\nu_3}\}$ with masses (eigenvalues) $m_1 < m_2<m_3$. Let $H$ be the Hamiltonian that describes neutrinos. The mass eigenstates satisfy
\begin{equation}
    H\ket{\nu_k}=E_k\ket{\nu_k} \quad (k=1,2,3)
\end{equation}
where $E_k =\sqrt{p^2 c^2+m_k^2 c^4}$, $p$ is the momentum of the neutrino, and $c$ is the velocity of light. 

These two basis states are related by a unitary matrix called the Pontecorvo–Maki–Nakagawa–Sakata matrix $U_\mathrm{PMNS}=(\langle \nu_\alpha|\nu_j\rangle)$ \cite{P1958,MNS1962,bib:pmns} as 
\begin{equation}
    \begin{pmatrix}
     \nu_e\\
     \nu_\nu\\
     \nu_\tau\\
     \nu_X
    \end{pmatrix}= U_\mathrm{PMNS}
     \begin{pmatrix}
     \nu_1\\
     \nu_2\\
     \nu_3\\
     \nu_4
    \end{pmatrix}
\end{equation}
where
\begin{equation}
    U_\mathrm{PMNS}
     = \begin{pmatrix}
 0.8255 & 0.5445 & -0.142+0.0434 i & 0 \\
 -0.2709+0.02739 i & 0.6057\, +0.0181 i & 0.7475 & 0 \\
 0.4938\, +0.0237 i & -0.5798+0.0157 i & 0.6475 & 0 \\
 0 & 0 & 0 & 1 
\end{pmatrix}.\label{eq:pmns}
\end{equation}
Here fictitious neutrinos $\nu_X$ and $\nu_4$ are introduced so that this system can be simulated with a two-qubit system. $\nu_X$ and $\nu_4$ are decoupled from the physical neutrinos and have no physical significance. This fourth neutrino may be utilized in a theory with an exotic neutrino which is yet to be discovered. 

We closely follow \cite{bib:neutrinoq} in the following. We keep the CP-violating phase $\delta_\mathrm{CP}$ in $U_\mathrm{PMNS}$ while \cite{bib:neutrinoq} ignored this phase. The phase was taken into account in \cite{bib:neutrinophase}, which also generalizes the simulation to arbitrarily many neutrino species. We employ parameters announced in November 2022 \cite{bib:pmns} in Eq.~(\ref{eq:pmns}). 

Suppose $\ket{\nu_\mu}=(0,1,0,0)^t$ is created at $(x,t) = (0,0)$. The probability of detecting $\ket{\nu_\alpha}$ $(\alpha=e, \mu, \tau)$ at $t>0$
is 
\begin{eqnarray}
     p_{\alpha}(t)
     &=&|\bra{\nu_\alpha} \nu_\mu(t)\rangle|^2
     = |\bra{\nu_\alpha} e^{-i H t/\hbar}|\nu_\mu\rangle|^2\nonumber\\
     &=&  \left|\sum_{k=1}^4 \bra{\nu_\alpha}\nu_k \rangle  e^{-i E_k t/\hbar}\bra{\nu_k}\nu_\mu\rangle \right|^2
     \nonumber\\
&=&|\bra{ \nu_\alpha} U_\mathrm{PMNS}\ \mathrm{diag}(e^{-i E_1 t/\hbar},  e^{-i E_2 t/\hbar},e^{-i E_3 t/\hbar},e^{-i \phi})\ U_\mathrm{PMNS}^\dag \ket{\nu_\mu}|^2,
\end{eqnarray}
where $\phi$ is an unphysical phase.

Let us simulate this system with a quantum computer. There are three steps to take.
\begin{itemize}
    \item Express $\ket{\nu_\mu}=\ket{01}$ in terms of the mass eigenstates $\ket{\nu_k}$, which is done by applying $U_\mathrm{PMNS}^\dag$ on $\ket{\nu_\mu}$.
    \item Apply the time-evolution operator $V(t)
    = e^{-iH t/\hbar}$ on this state to find $\ket{\nu_\mu(t)}= V(t) U_\mathrm{PMNS}^\dag \ket{\nu_\mu}$. $V(t)$ is decomposed into two one-qubit gates as $V(t)=S_1(t) \otimes S_2(t)$,
    whose explicit forms are given below.
    \item Measure $\ket{\nu_\mu(t)}$ in the basis $\{\ket{\nu_e}, \ket{\nu_\mu}, \ket{\nu_\tau}\}$. This is done by applying $U_\mathrm{PMNS}$ on $\ket{\nu_\mu(t)}$ and measure the output with binary basis.
\end{itemize}

Let us analyze these steps in depth. We prepare the initial state $\ket{\nu_\mu}=\ket{01}$ in the $\{\ket{\nu_k}\}$ basis as 
\begin{equation}
    U_\mathrm{PMNS}^\dag \ket{10}
    = u_{\mu 1}^*\ket{\nu_1} + u_{\mu 2}^*\ket{\nu_2}+u_{\nu 3}^*\ket{\nu_3},
\end{equation}
where $u_{\mu k}$ are the matrix elements of
$U_\mathrm{PMNS}$. 

The time-evolution operator $V(t)$ is diagonal in this basis and takes the form
\begin{equation}
    V(t)=\mathrm{diag}(e^{-i E_1 t/\hbar}, e^{-i E_2 t/\hbar},e^{-i E_3 t/\hbar},e^{-i \phi}).
\end{equation}
Since the overall phase has no physical significance, we may factor out $e^{-i E_1 t/\hbar}$ so that
\begin{equation}
    V(t)=\mathrm{diag}(1, e^{-i E_{21} t/\hbar},e^{-i E_{31} t/\hbar},e^{-i \phi'})
    =S_1(t) \otimes S_2(t),
\end{equation}
where $E_{21}=E_2-E_1$, $E_{31}=E_3-E_1$ and $\phi'$ is another unphysical phase. Here
$$
S_1(t) = \begin{pmatrix}
 1&0\\
 0&e^{-i E_{31} t/\hbar}
\end{pmatrix}, \quad S_2(t) = \begin{pmatrix}
 1&0\\
 0&e^{-i E_{21} t/\hbar}
\end{pmatrix},
$$
where we took advantage of the fact that $\phi'$ has no physical meaning. Now we have $\ket{\nu_\mu(t)}$, the state of neutrino at $t$, which was $\ket{\mu_\nu}$ at $t=0$, as
\begin{equation}
    \ket{\nu_\mu(t)} = (S_1(t) \otimes S_2(t))U_\mathrm{PMNS}^\dag \ket{01}.
\end{equation}
Since the neutrino masses are very small and the velocity is very close to the speed of light $c$, we may approximate $E_k=\sqrt{p^2c^2+m_k^2 c^4}$ as $E_k \simeq pc + m_k^2 c^3/2p$. Then 
\begin{equation}
E_{k1} \simeq \frac{\Delta m_{k1}^2 c^3}{2p}
\quad (k=2,3)
\end{equation}
where $\Delta m_{k1}^2=m_k^2-m_1^2$. We employ $\Delta m_{21}^2= 7.39 \times 10^{-5} \mathrm{eV}^2$ and $\Delta m_{31}^2= 2.45 \times 10^{-3} \mathrm{eV}^2$ in our analysis~\cite{bib:pdg}.

By approximating $E \simeq pc$ and $L \simeq ct$, we have
\begin{equation}
    e^{-i E_{k1}t/\hbar} \simeq 
    \exp\left(- i \frac{\Delta m_{k1}^2 c^3 }{2\hbar}\frac{L}{E}\right).
\end{equation}
The exponent is expressed numerically with physical units as
$$
\frac{\Delta m_{k1}^2 c^3 }{2\hbar} \frac{L}{E}
\simeq 2.534\, \Delta m_{k1}^2\, [\mathrm{eV}^2]\times \frac{L\,[\mathrm{km}]}{E\,[\mathrm{GeV}]}.
$$

\begin{figure}
\centering
\subfloat[]{\includegraphics[width=0.75\columnwidth]{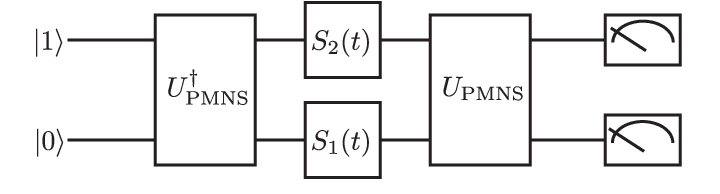}}
\\
\subfloat[]{\includegraphics[width=0.8\columnwidth]{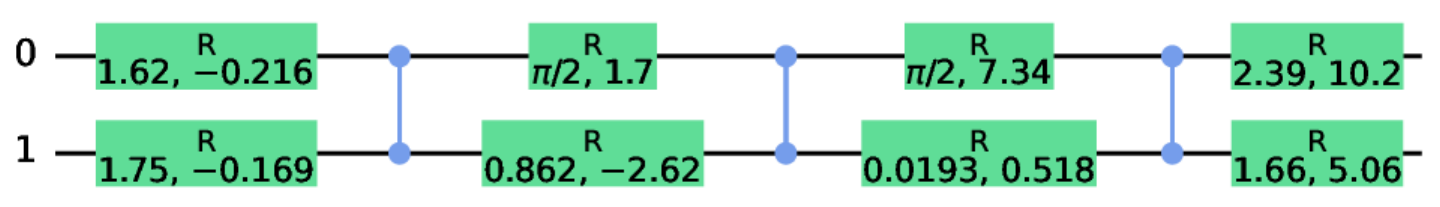}}
\\
\subfloat[]{\includegraphics[width=0.9\columnwidth]{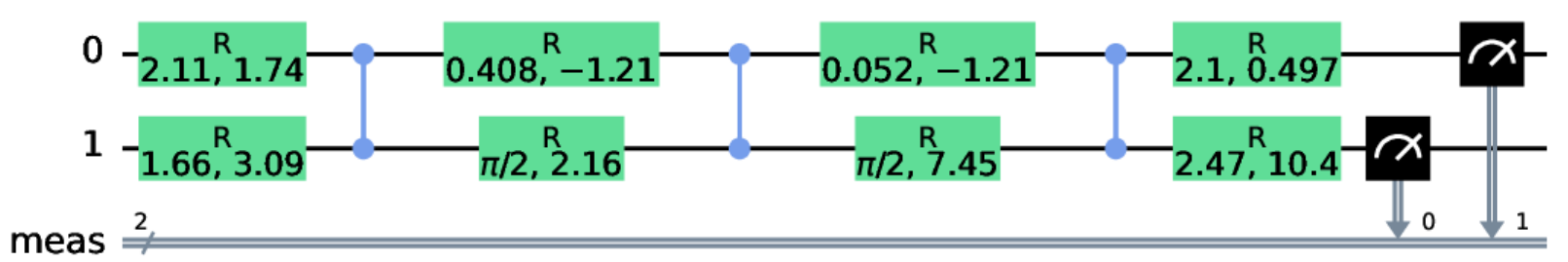}}
\caption{(a) Quantum circuit to simulate neutrino oscillation among 3 generations. The input state is $\ket{01}=\ket{\nu_\mu}$. (b) Implementation of $U_\mathrm{PMNS}$ using our native gates. (c) Circuit to simulate neutrino oscillations, transpiled to our native gates at $L/E=4000$. The number of gates is reduced via circuit optimization during the transpilation. Note that in (b) and (c) the order of qubits is reversed according to the Qiskit convention. The second parameter of $R$ is defined $\text{mod}~ 2\pi.$ \label{fig:oscillation4}}
\end{figure}

These steps are implemented with a quantum computer with different $t$, namely different $L/E$ and the results are compared with theoretical prediction. Figure~\ref{fig:oscillation4}~(a) shows the quantum circuit for this scheme while (b) shows the gate decomposition of $U_\mathrm{PMNS}$. The overall gate decomposition depends on $L/E$. Figure~\ref{fig:oscillation4}~(c) shows the transpiled gate decomposition for $L/E=4000$\,[km/GeV].

\begin{figure}
    \centering
    \includegraphics[width=0.68\columnwidth]{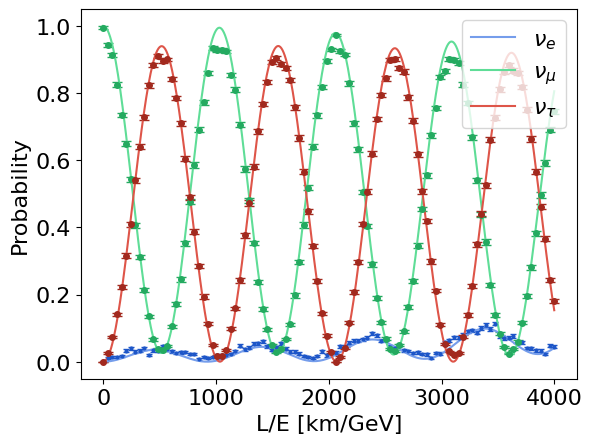}
    \caption{Predicted measurement probabilities of neutrino species (solid curves) and measured probabilities (markers) as functions of $L/E~[\mathrm{km/GeV}]$, where $\nu_\mu$ is created at $L=0$ at $t=0$. We used 5000 shots per data point and applied readout error mitigation.}
    \label{fig:3gens}
\end{figure}

Figure \ref{fig:3gens} shows the theoretical curves predicting the detection probabilities of three neutrinos and our quantum computer output as functions of $L/E$.
Probabilities oscillate as a function of $L/E$ while the sum of probabilities is always 1. This oscillation is possible only when different types of neutrinos have different masses ($\Delta m_{k1}\neq 0$). Thus neutrino oscillation is a smoking gun of massive neutrinos.\footnote{The ``standard model'' of particle physics had no neutrino mass terms for many decades.} Neutrino oscillation was the subject of the Nobel Prize in Physics in 2015 \cite{ref:nobelneutrino}.

\subsection{Estimation of the Jones polynomials}

Knots, links and braids are fascinating subjects of topology. An unexpected encounter between mathematics (topology) and physics (statistical physics) was discovered by Vaughan Jones in 1984. He discovered knot and link invariants, later called the Jones polynomials, that characterize oriented knots and links. His work was inspired by statistical mechanics of generalized spin models defined on a lattice. Jones was awarded the Fields medal in 1990 for his achievements including the discovery of the Jones polynomials.

Estimation of the Jones polynomials by employing a quantum computer was proposed in \cite{ref:jonesqc1,ref:jonesqc2} and demonstrated with an NMR quantum computer \cite{ref:jonesnmr}. We reproduce here the results of  
\cite{ref:jonesnmr} using our superconducting quantum computer.

\begin{figure}
\centering
\includegraphics[width=8cm]{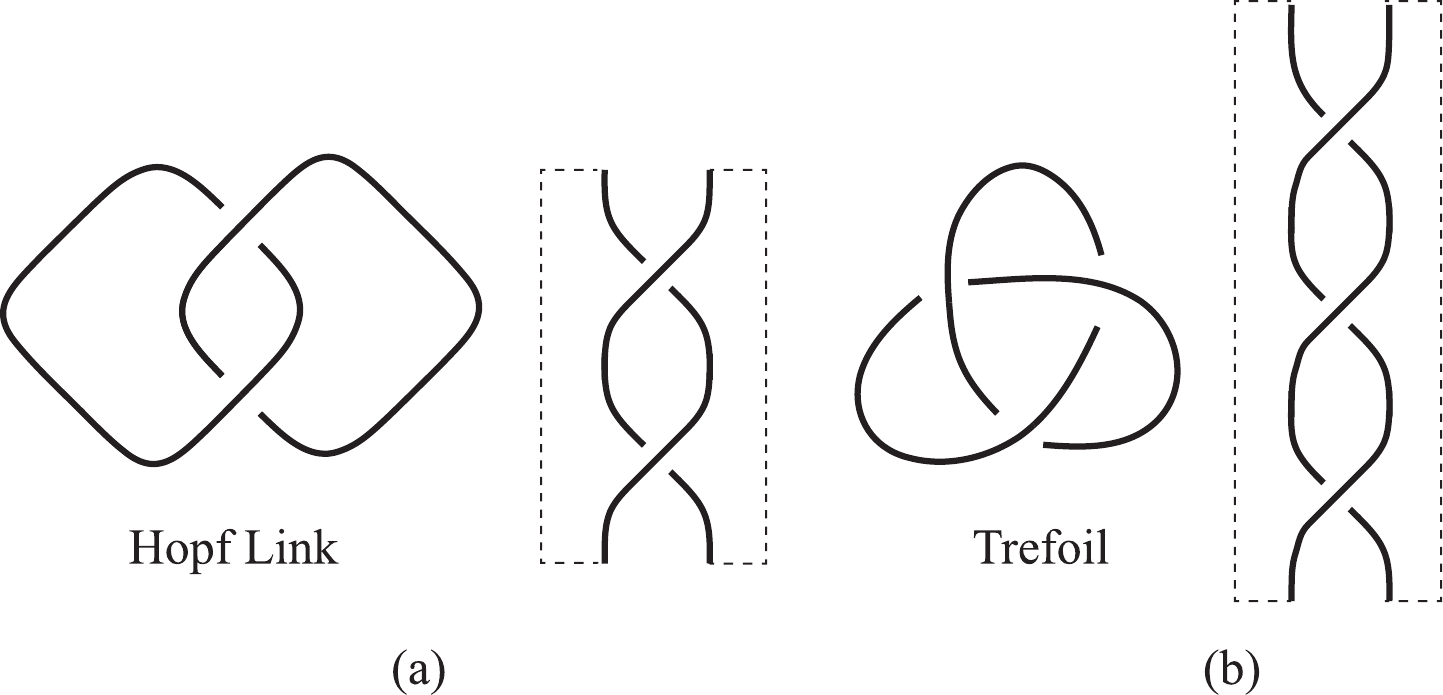}
\caption{(a) Hopf link and its corresponding braid. (b) Trefoil knot and its corresponding braid. Braids are closed with dotted lines to form the link and the knot.}
\label{fig:braids_knots}
\end{figure}

\begin{figure}
    \centering
    \includegraphics[width=5cm]{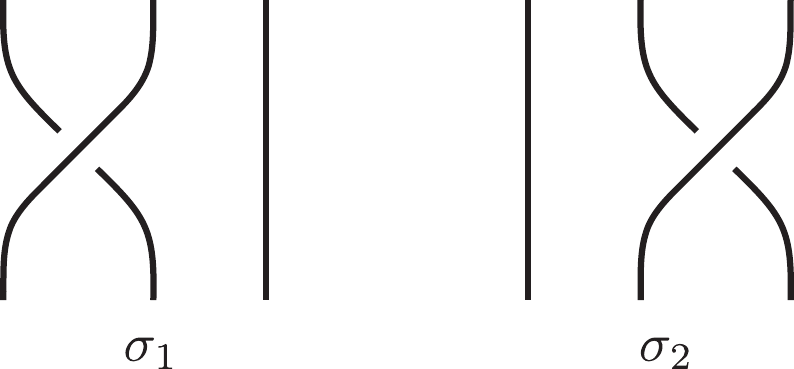}
    \caption{Generators $\sigma_1, \sigma_2$ of three-strand braids.}
    \label{fig:braid_generators}
\end{figure}

We first introduce a braid $b$ associated with a link $L$. A link is an embedding of a set of loops in $\mathbb{R}^3$ or the 3-dimensional sphere $S^3$. If a link is made of one component, it is called a knot $K$. Let $(x,y,t)$ be a coordinate of 3-dimensional space-time. A braid $b$ is a set of strings that connect $n$ points $(0,0,0), (1,0,0), \ldots (n,0,0)$ on $\mathbb{R}^2$ at $t=0$ and $(0,0,1), (1,0,1), \ldots (n,0,1)$ at $t=1$ without intersecting or going backwards in time. Alexander's theorem claims that every knot and link can be expressed as a braid whose end points are closed, see Fig.~\ref{fig:braids_knots}.
The set of braids with $n$ strands has a group structure called the Artin group $B_n$, whose generators are denoted as $\sigma_k$. The generator $\sigma_k$ twists the $k$th strand and $(k+1)$st strand as shown in Fig.~\ref{fig:braid_generators} for $n=3$. The inverse $\sigma_k^{-1}$ twists them in the opposite direction. The set of generators satisfy the following relations:
\begin{equation}\label{eq:braidgen}
\begin{array}{cl}
    \sigma_k \sigma_k ^{-1}=1,&k =1,2, \ldots, n-1\\
    \sigma_k \sigma_{k+1} \sigma_k=\sigma_{k+1}\sigma_k \sigma_{k+1},&
    k=1, 2, \ldots, n-2\\
    \sigma_j \sigma_k=\sigma_k \sigma_j,&
    |j-k|\geq 2.
    \end{array}
\end{equation}
An arbitrary braid $b$ can be expressed in terms of successive applications of these generators and their inverses as
\begin{equation}\label{eq:braidword}
    \sigma_{j_p}^{s_p} \sigma_{j_{p-1}}^{s_{p-1}} \ldots \sigma_{j_2}^{s_2} \sigma_{j_1}^{s_1},
\end{equation}
where $s_k \in \{1, -1\}$, $j_k \in \{1, 2, \ldots, n-1\}$ and $\sigma_{j_1}^{s_1}$ is applied first and $\sigma_{j_p}^{s_p}$ last. This is called the braid word of $b$. Braid words are not unique and there are infinitely many braid words corresponding to the same braid. 

\begin{figure}
    \centering
    \includegraphics[width=7cm]{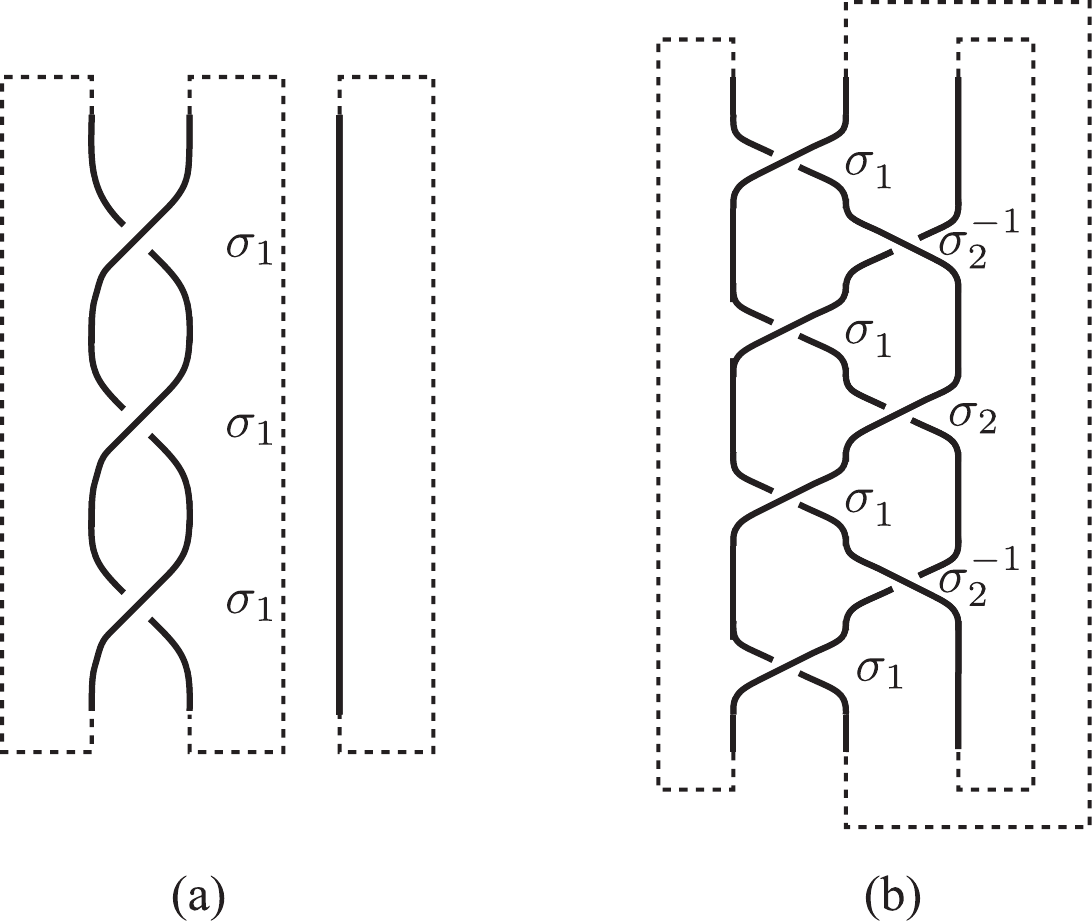}
    \caption{(a) Simplest representation $\sigma_1^3$ of the trefoil knot. (b) Equivalent but more knotted representation of the trefoil knot.}
    \label{fig:trefoiltwist}
\end{figure}

In the following, we are concerned with three-strand braids, namely $n=3$. It has two generators $\sigma_1$ and $\sigma_2$ as shown in Fig.~\ref{fig:braid_generators}.
One possible braid word of the trefoil knot (Fig.~\ref{fig:braids_knots} (b)) is $\sigma_1^3$. Closure of the three-strand braid $\sigma_1^3$ results in the trefoil and the trivial knot, as shown in Fig.~\ref{fig:trefoiltwist}~(a).

It is possible to ``unwind'' crossing of a braid diagram by introducing the generators $\{U_i\}$ of the Temperley-Lieb algebra $\text{TL}_n$. There are two generators $U_1, U_2$ for $n=3$ and they satisfy
\begin{equation}\label{eq:tlgen}
    U_i^2=\delta U_i,\ U_1 U_2 U_1= U_1,\ U_2 U_1 U_2= U_2 
\end{equation}
where $\delta =-A^2-A^{-2}$ with $A=e^{i \theta}$ a complex number with unit modulus.
The map $\rho: B_3 \to \mathrm{GL}_2(\mathbb{C})$
defined as
\begin{equation}
    \rho(\sigma_i)= A I + A^{-1} U_i
\end{equation}
is a representation of $B_3$. We take the simplest braid word of the trefoil $\sigma_1^3$ and its representation $\rho(\sigma_1^3)$ here. Note that the relations of the braid generators (\ref{eq:braidgen}) are satisfied if $U_i$ satisfies Eq.~(\ref{eq:tlgen}). The representation $\rho$ is unitary whenever $U_i$ is a real symmetric matrix satisfying $\delta^2 \geq 1$. The last condition is satisfied if
$$
\theta \in [0,\pi/6] \sqcup [\pi/3, 2\pi/3] \sqcup [5\pi/6, 7\pi/6] \sqcup [4\pi/3,5\pi/3] \sqcup [11\pi/6, 2\pi].
$$
Explicitly, the generators $U_i$ are given as
\begin{equation}
    U_1= \begin{pmatrix}
     \delta&0\\
     0&0
    \end{pmatrix}\quad \mbox{and}\quad
      U_2= \begin{pmatrix}
     \delta^{-1}&\sqrt{1-\delta^{-2}}\\
     \sqrt{1-\delta^{-2}}&\delta -\delta^{-1}
    \end{pmatrix}.
\end{equation}
We need to define $w(b) \in \mathbb{Z}$ called the writhe of a braid $b$ before we introduce the Kauffman bracket and the Jones polynomial. Suppose a braid word of $b$ is given as Eq.~(\ref{eq:braidword}). Then the writhe of $b$ is defined as the sum of the exponents,
\begin{equation}
     w(b) = \sum_{i=1}^p s_i.
\end{equation} 
For the Hopf link and the trefoil knot we find $w(\mbox{Hopf link})=2$ and $w(\mbox{trefoil})=3$, respectively.

The Kauffman bracket of the trefoil is obtained as 
\begin{equation}\label{eq:trefoiljones}
    \langle \mbox{trefoil} \rangle =\frac{\langle \bar{b} \rangle}{\delta}= \frac{1}{\delta}(\mathrm{tr} \rho(\sigma_1^3)+A^{w(b)}(\delta^2-2)) =-A^5 - A^{-3} + A^{-7},
\end{equation}
where $b$ is the braid word given in Fig.~\ref{fig:braids_knots} (b) and $\bar{\ }$ stands for the closure of $b$. Note that $\bar{b}$ is made of the trefoil knot {\it and} the trivial knot. The factor $1/\delta$ in Eq.~(\ref{eq:trefoiljones}) removes the contribution of the trivial knot.

The Kauffman bracket of the Hopf link is obtained in a similar way as
\begin{equation}
\langle \mbox{Hopf link} \rangle = \frac{1}{\delta}
(\mathrm{tr}\rho(\sigma_1^2) + A^2(\delta^2-2))
=-A^4-A^{-4}.
\end{equation}

The Kauffman bracket is invariant under the Reidemeister moves II and III but not under the Reidemeister move I, and hence cannot be a knot invariant. The Jones polynomial is obtained by multiplying the Kauffman bracket with $(-A^3)^{- w(b)}$ to make it invariant under all three Reidemeister moves. For the trefoil knot, we obtain the Jones polynomial
\begin{equation}
    V_\mathrm{trefoil}(A)= (-A^3)^{-3}(-A^5 - A^{-3} + A^{-7})
    = A^{-4}+ A^{-12}- A^{-16}. 
\end{equation}
It is common to introduce $t=A^{-4}$ so that
\begin{equation}
    V_\mathrm{trefoil}(t) = - t^4 + t^3 + t.
\end{equation}
The Jones polynomial of the Hopf link is
\begin{equation}
    V_\mathrm{Hopf\ link} = -A^{-10}-A^{-2} = -\sqrt{t}(1+t^2).
\end{equation}
The Jones polynomial is a Laurent polynomial in $\sqrt{t}$ in general.

\begin{figure}
\centering
\includegraphics[width=0.6\columnwidth]{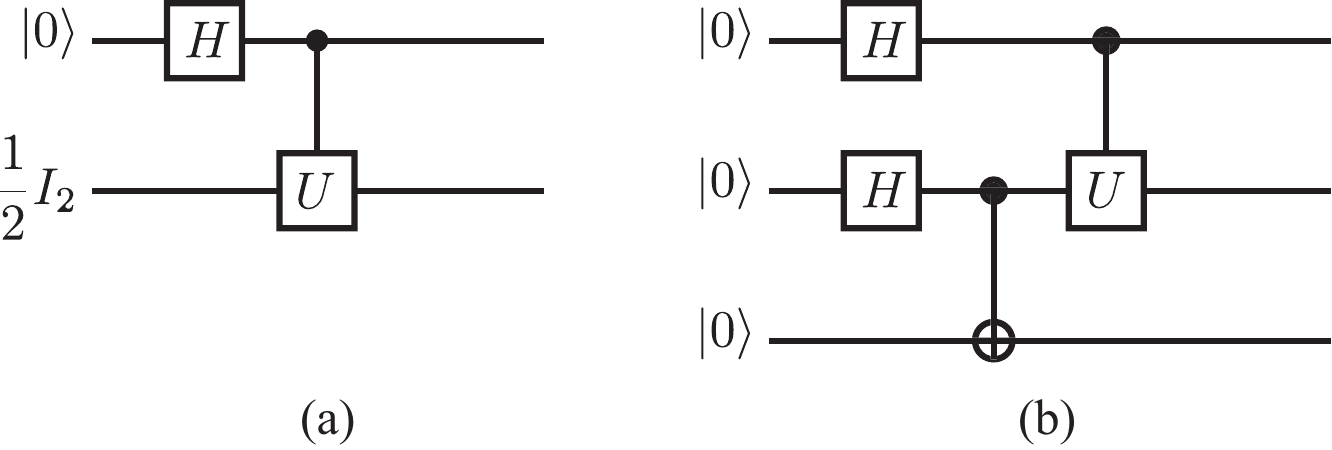}
\\
\includegraphics[width=1.0\columnwidth]{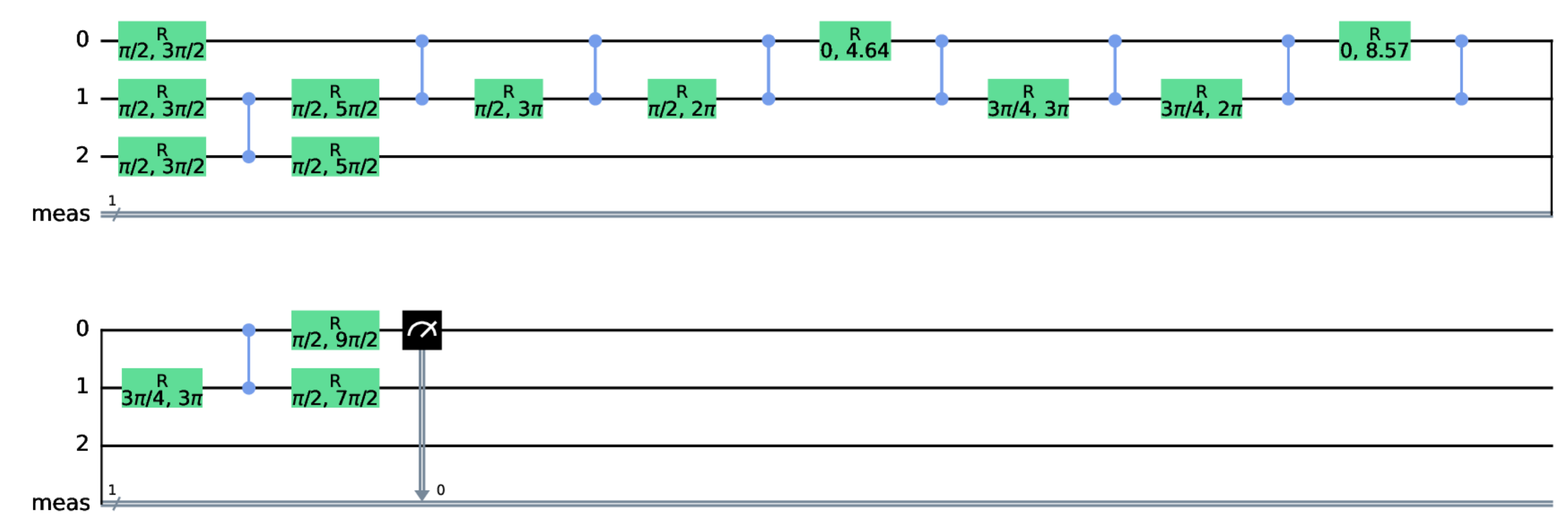}
\\(c)
\caption{(a) Quantum circuit to estimate the Jones polynomial with input $\ket{0}\bra{0} \otimes I_2/2$. (b) Quantum circuit to estimate the Jones polynomial using ancilla qubit to synthesize the mixed state. The input state is $\ket{000}$. To estimate the Jones polynomial, the expectation values of $X_1$ and $Y_1$ are estimated for this state. (c) Circuit for the trefoil knot, appended with a $H$ gate on the first qubit for measuring $E(X_1)$ and transpiled to our native gates and connectivity at $\theta=\pi/6$. The second parameter of $R$ is defined $\text{mod}~ 2\pi.$}
\label{fig:jonescircuit}
\end{figure}

Let us use a quantum computer to estimate the trace $\mathrm{tr} \rho(\sigma_1^k)$ in the Kauffman bracket. Consider the quantum circuit Fig.~\ref{fig:jonescircuit} (a) with $U= \rho(\sigma_1^k)$ and the input state
$$
\rho_0 = \ket{0}\bra{0} \otimes \frac{1}{2}I_2
=\frac{1}{2}\begin{pmatrix}
 I_2&0\\
 0&0
\end{pmatrix},
$$
where $I_2/2$ is the maximally mixed state. The state after quantum circuit is applied is
$$
\rho_1=\begin{pmatrix}
 I_2&0\\
 0&U
\end{pmatrix}\frac{1}{4}\begin{pmatrix}
 I_2&I_2\\
 I_2&I_2
\end{pmatrix} \begin{pmatrix}
 I_2&0\\
 0&U^\dag
\end{pmatrix} =\frac{1}{4}\begin{pmatrix}
 I_2&U^\dag \\
 U&I_2
\end{pmatrix}.
$$
The expectation value of $X_1$ with respect to $\rho_1$ is
\begin{equation}
     E(X_1) = \mathrm{tr}(X_1 \rho_1)
    = \frac{1}{2} \mathrm{Re\ tr}\,U
\end{equation}
while the expectation value of $Y_1$ is 
\begin{equation}
    E(Y_1) = \mathrm{tr}(Y_1 \rho_1)
    = \frac{1}{2} \mathrm{Im\ tr}\,U.
\end{equation}
Hence, the $\mathrm{tr}\, U$ is found from estimating these two expectation values.

The above scheme fits well with NMR quantum computer, in which the system is in a maximally mixed state with a good approximation. A superconducting quantum computer is ideally in a pure state and the above scheme cannot be applicable in its original form. We use purification to ``synthesize'' a uniformly mixed state from a pure state for this purpose. Let us consider the circuit in Fig.~\ref{fig:jonescircuit} (b). The bottom two qubits are in the Bell state
$$
\ket{\Phi_+} =\frac{1}{\sqrt{2}}(\ket{00}+\ket{11})
$$
after application of the Hadamard gate and the CNOT gate. The middle qubit is in a maximally mixed state if the bottom qubit is ignored (i.e. partially traced out). We have already seen this in section 4.2.2. Observe that
$$
\mathrm{tr}_2 \ket{\Phi_+}\bra{\Phi_+}=\frac{1}{2} I_2.
$$
Then the output of the top two qubits is the same as that of Fig.~\ref{fig:jonescircuit} (a). 

Suppose the quantum circuit Fig.~\ref{fig:jonescircuit} (b) is applied on $\ket{000}$. Then the output state is
\begin{equation}
    \ket{\Psi}=\frac{1}{2}(\ket{000} + \ket{011} +\ket{1}(U\ket{0})\ket{0} +\ket{1}(U\ket{1})\ket{1}).
\end{equation}
We estimate the expectation value of $X_1$ with respect to $\ket{\Phi}$ to get
\begin{equation}
    E(X_1)= \bra{\Phi} X_1 \ket{\Phi} =
    \frac{1}{4}(\mathrm{tr}\,U+ \mathrm{tr}\,U^\dag)=\frac{1}{2}\mathrm{Re}(\mathrm{tr}\, U),
\end{equation}
similarly for $Y_1$ we get 
\begin{equation}
    E(Y_1)=\bra{\Phi} Y_1 \ket{\Phi} =
    \frac{i}{4}(- \mathrm{tr}\,U+  \mathrm{tr}\,U^\dag)=
    \frac{1}{2}\mathrm{Im}(\mathrm{tr}\, U),
\end{equation}
from which we estimate $\mathrm{tr}\,U$. Figure~\ref{fig:jonescircuit}~(c) shows the transpiled circuit to estimate $E(X_1)$ for $\theta=\pi/6$.

\begin{figure}
\centering

\subfloat[]{
\includegraphics[width=0.45\columnwidth]{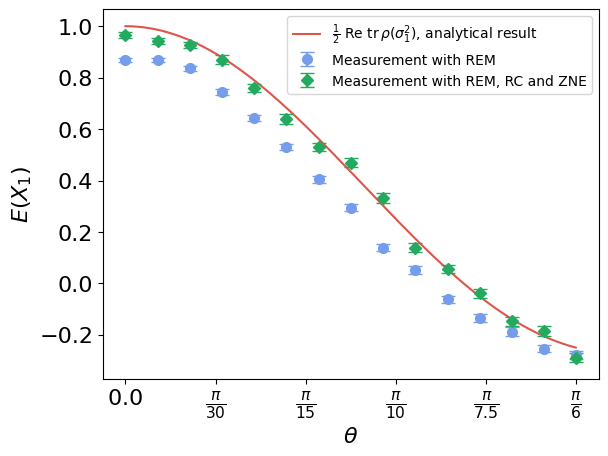}}
\subfloat[]{
\includegraphics[width=0.45\columnwidth]{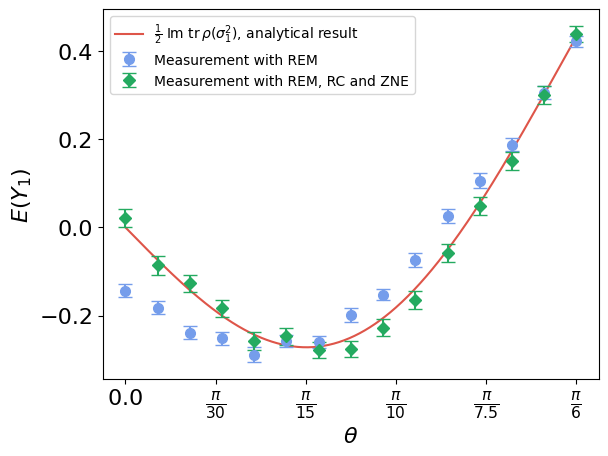}}
\\
\subfloat[]{
\includegraphics[width=0.45\columnwidth]{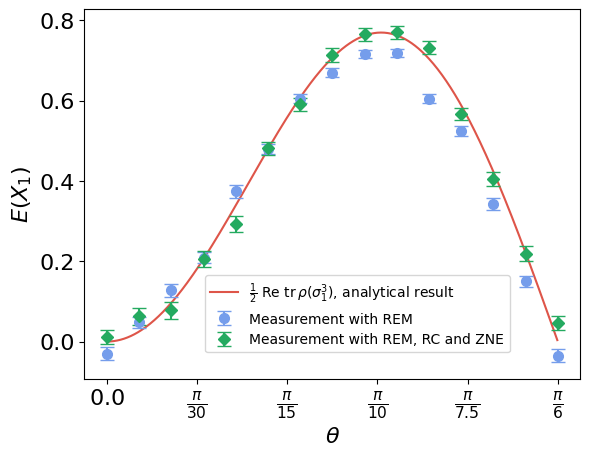}}
\subfloat[]{
\includegraphics[width=0.45\columnwidth]{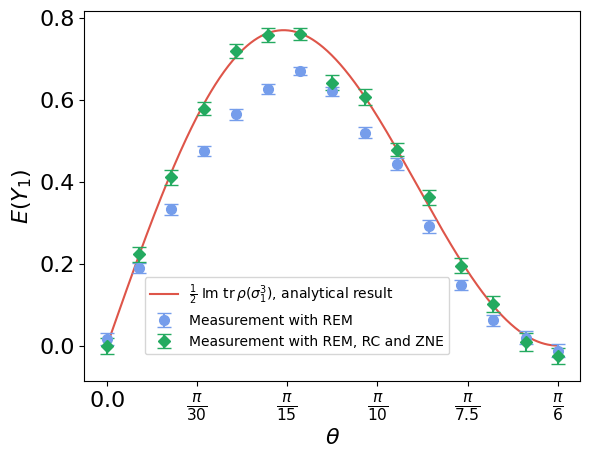}}

\caption{Trace in the Jones polynomial of the Hopf link and the trefoil knot. (a) Real part and (b) imaginary part of the trace in the Hopf link, estimated as the expectation values of $X$ and $Y$ on the first qubit, respectively. Similarly, (c) and (d) show the real and the imaginary parts of the trace for the trefoil knot. Solid curves are theoretical results while the markers show experimental results evaluated on our quantum computer. Circles are from readout error mitigated execution and diamonds include also randomized compiling (RC) and zero-noise extrapolation (ZNE).}
\label{fig:trefoiljp}
\end{figure}

Figure \ref{fig:trefoiljp} shows the real and the imaginary parts of $\mathrm{tr}\, \rho(\sigma_1^k)$ obtained using $\ket{\Phi}$ with $k=2$ for the Hopf link while $k=3$ for the trefoil knot. The readout error mitigated results are marked by the blue circles.

When quantum circuits get deeper, consisting of several layers of gates, errors that happen in the bulk of the circuits start to accumulate and can significantly affect the results. This is the case for the transpiled circuits that we are considering here, see Fig.~\ref{fig:jonescircuit} (c), and we thus decided to implement also error suppression and additional error mitigation techniques, namely randomized compiling and zero-noise extrapolation. The mitigated expectation values are shown with green diamonds and, in general, are indeed closer to the ideal noiseless results (red). In more detail, we generate 30 different randomized compilations of the original circuit and measure each 20,000 times. The higher shot count is employed to combat the increased variance resulting from zero-noise extrapolation, where we scale the noise by factors of 3 and 5 using global folding. Polynomial fitting is used to extrapolate to the limit of zero noise.  

The Kauffman and the Jones polynomials may be obtained in many different ways. All of them are easy if the trefoil is represented as in Fig. \ref{fig:trefoiltwist}~(a), for example. But it will be more demanding if the representation is more knotted as in Fig. \ref{fig:trefoiltwist}~(b).
A typical classical evaluation of these polynomials involves a sum over ``states'' obtained by splitting each crossing in two different ways. There are $2^m$ states if there are $m$ crossings and the task is exponentially hard as $m$ increases. In contrast, in quantum computing, a controlled unitary gate is assigned to each crossing, i.e., the braid group generator, which requires merely $m$ controlled unitary gates.

\subsection{Embedding Techniques for Quantum Chemistry}
\noindent
Embedding techniques are theoretical frameworks used in quantum chemistry and condensed matter physics to study the electronic structure of strongly correlated materials. These methods are crucial and are particularly valuable for systems where the effects of electron-electron interactions are significant and where traditional methods such as density functional theory (DFT) \cite{hohenberg_1964,kohn_1965} fail to provide accurate descriptions. They represent a powerful tool to study the electronic properties of materials such as transition-metal-oxides and rare-earth compounds. Embedding techniques have been successfully applied to further the understanding of complex phenomena such as metal-insulator transitions, magnetically ordered states and high-temperature superconductivity. 

\begin{figure}
\begin{center}
\includegraphics[width=0.75\linewidth]{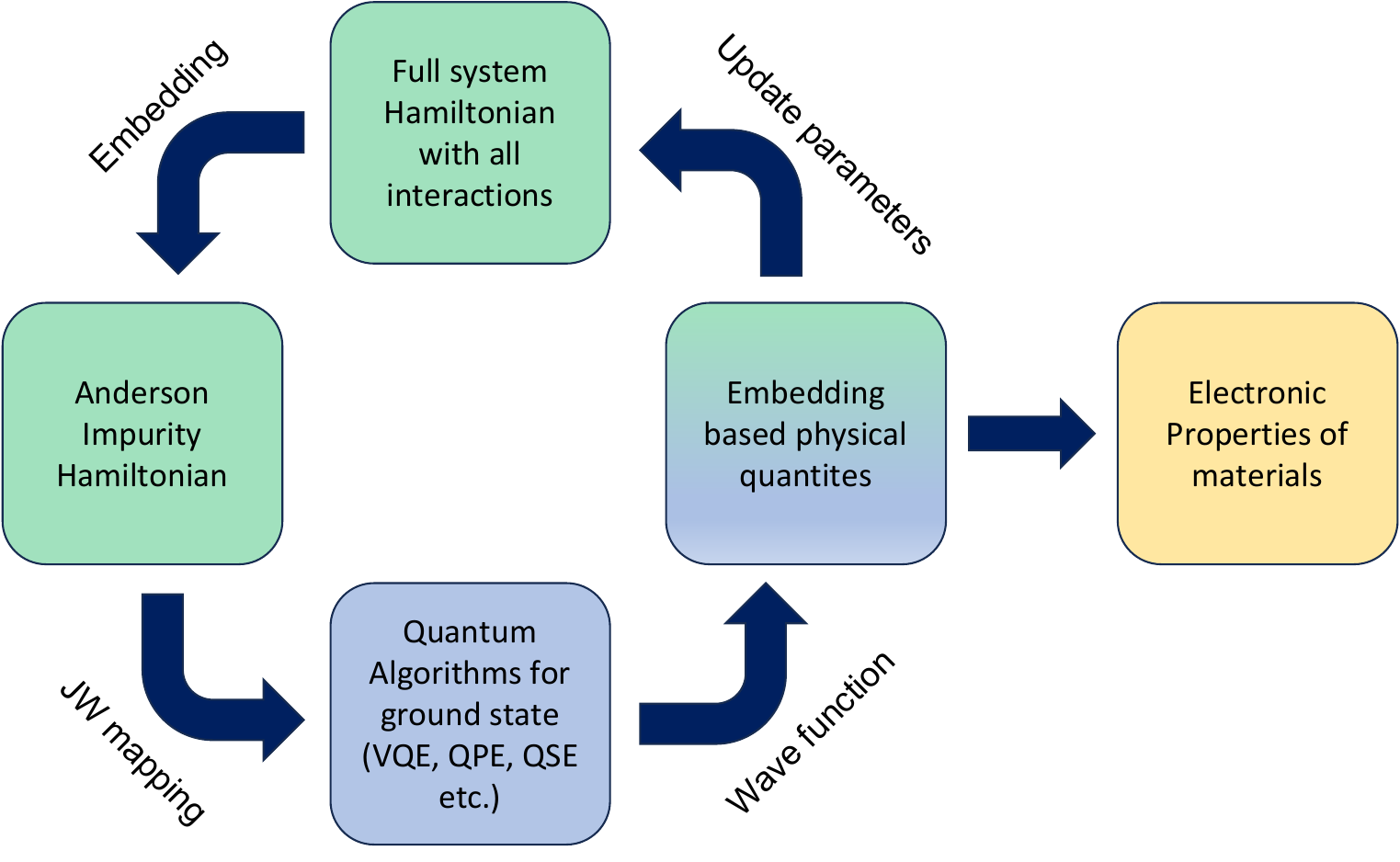}
\caption{Schematic diagram of the embedding loop showcasing the part of the loop that is to be performed on the quantum computer.}
\label{fig:dmft_loop}
\end{center}
\end{figure}

The central idea of these techniques is to map a complex quantum many-body problem to a self-consistent Anderson impurity model (AIM) \cite{anderson_1961}, which consists of a strongly correlated subsystem (impurity) and a weakly correlated or non-correlated subsystem (bath). The impurity is treated more accurately using a method capable of handling strong correlations, while the bath is treated at a lower level of theory, often using mean-field approximations. These descriptions of the impurity and the bath are combined in a self-consistent loop over either single particle (i.e. density matrix embedding theory (DMET) \cite{knizia_2012,wouters_2016} and rotationally invariant slave-boson (RISB) techniques \cite{lechermann_2007,fresard_1992,ayral_2017}) or two particle quantities (i.e. self-energy embedding theory (SEET) \cite{zgid_2017,lan_2017}, dynamical mean-field theory (DMFT) \cite{georges_1996,kotliar_2006,metzner_1989} and its cluster extensions \cite{kotliar_2006,park_2008}) in order to provide a more accurate description of the entire system. Figure~\ref{fig:dmft_loop} is a schematic of the self-consistent loop in embedding based techniques.

Consider the electronic structure (ES) Hamiltonian in second-quantized form:
\begin{equation}
    \mathcal{H}_{\text{ES}} = \sum_{pq}h^{pq} a^\dagger_p a_q^{\phantom{\dagger}} + \sum_{pqrs}h^{pqrs} a^\dagger_p a^\dagger_q a_r^{\phantom{\dagger}} a_s^{\phantom{\dagger}}
\end{equation}
where $p,q,r,s$ are indices of a given basis set and include the respective spin indices $p \equiv p(\sigma)$. Further, $a^\dagger$ and $a$ are the fermionic creation and annihilation operators and $h^{pq}$, $h^{pqrs}$ are constants called the one- and two-electron integrals, respectively. Such a Hamiltonian is then mapped to an Anderson impurity-bath model given by:
\begin{align}
    \mathcal{H}_{\text{AIM}} = \sum_{k,\sigma} \varepsilon_k c_{k\sigma}^\dagger c_{k\sigma}^{\phantom{\dagger}} + \sum_\sigma (\varepsilon_d - \mu) d_\sigma^\dagger d_\sigma^{\phantom{\dagger}} + U n_{d\uparrow}^{\phantom{\dagger}} n_{d\downarrow}^{\phantom{\dagger}} + \sum_{k,\sigma} (V_k c_{k\sigma}^\dagger d_\sigma^{\phantom{\dagger}} + \text{H.c.}). \label{eq:aim}
\end{align}
where \(k\) represents the summation index for the bath operators, $\sigma$ labels the spin, \(d\) corresponds to the impurity operator with $n_{d\sigma} = d^\dagger_\sigma d_\sigma$ and \(c\) corresponds to the operators on the non-interacting bath, $\varepsilon_{d/c}$ correspond to the onsite-energy of the impurity and bath, respectively, with the chemical potential $\mu$. In spite of giving a simple description of the lattice problem, this model is by itself challenging to solve for state-of-the-art numerical techniques such as density matrix renormalization group (DMRG) \cite{white_1992,white_1993,schollwock_2005,schollwock_2011}, quantum Monte Carlo (QMC) \cite{blankenbecler_1981,gull_2011,foulkes_2001,al_2021} and others. It has been recently shown that a promising alternative approach for computing the ground states of such systems comes from using variational algorithms implemented on quantum devices \cite{wecker_2015,cerezo_2021} one of which, QAOA, has already been discussed in section \ref{sec:va}.

The variational quantum eigensolver (VQE) \cite{peruzzo_2014,mcclean_2016,cao_2019} is another quantum algorithm designed for finding the ground state energy, which is a fundamental quantity for quantum systems of interest in condensed matter physics and quantum chemistry. The basic idea behind VQE is to use a hybrid approach where the optimization of variational parameters performed on a classical computer is combined with measurements on the quantum computer to find an approximation to the ground state energy. The algorithm involves mainly a trial wavefunction in the form of a parameterized quantum circuit, chosen to represent a guess for the ground state of the quantum system. This circuit, controlled by a set of variational parameters, is executed on a quantum computer. Measurements are made on the final prepared state to calculate the expectation value of the Hamiltonian of the system. Then classical optimization algorithms are employed to tune the variational parameters in order to minimize the expectation value of the Hamiltonian. This process is repeated iteratively until the ground state energy is sufficiently minimized and the solution converges.

\begin{figure}
\begin{center}
\includegraphics[width=0.8\linewidth]{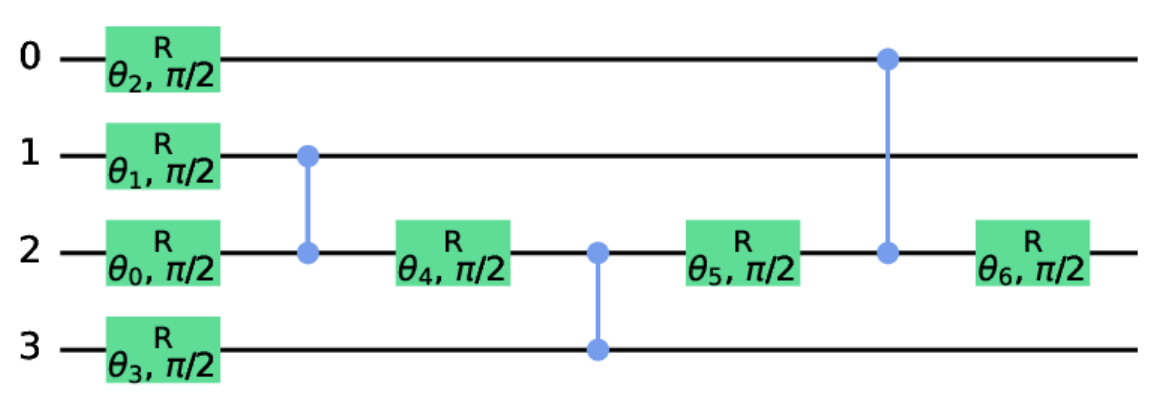}\\
(a)\\
\includegraphics[width=0.89\linewidth]{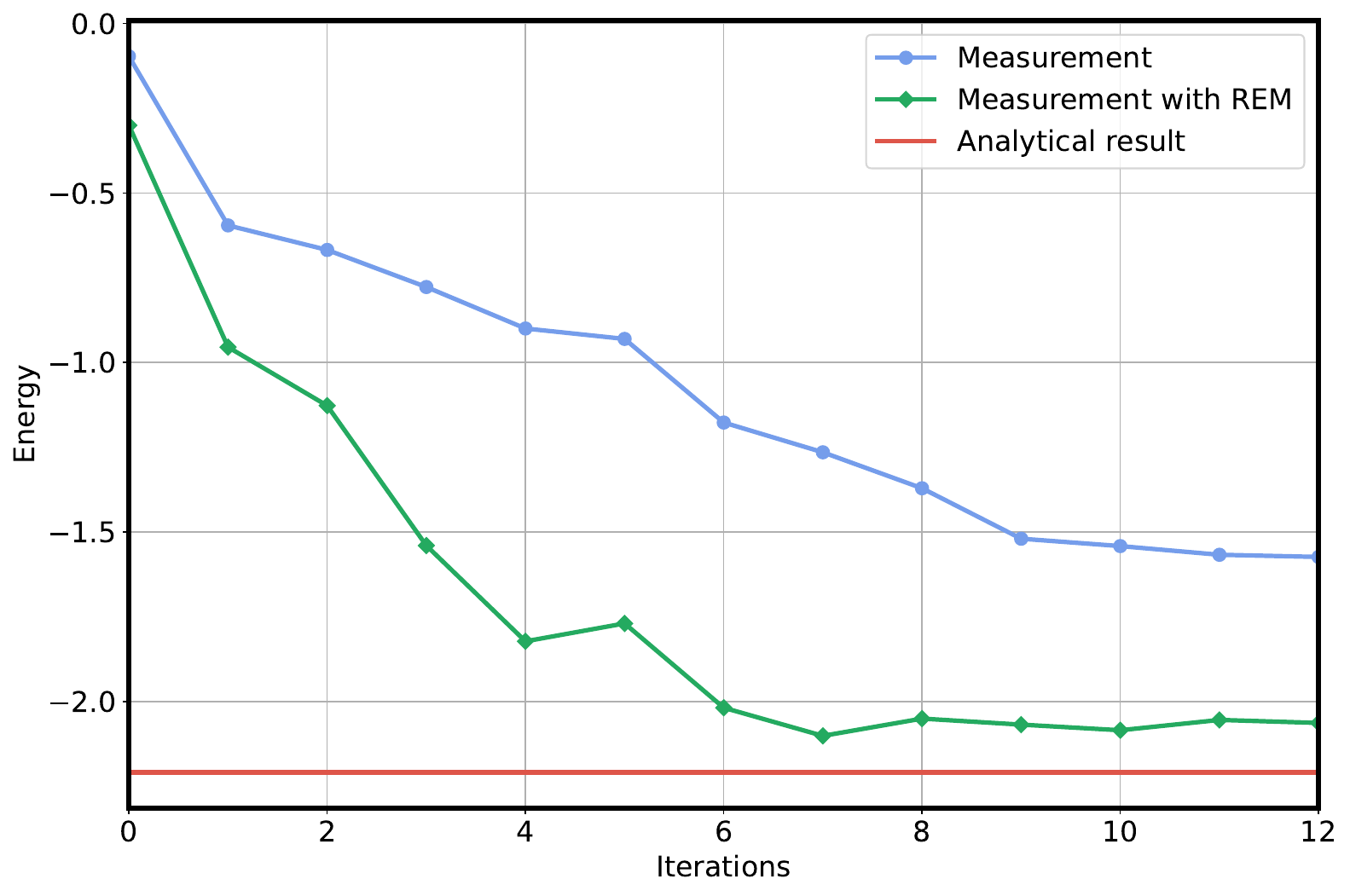}
\\
(b)
\caption{(a) Hardware efficient Ansatz used for the VQE. (b) Comparison of the VQE energy as a function of the iterations for convergence of an Anderson impurity model at $U = 2$ on our 5-qubit quantum computer with and without readout error mitigation. The exact energy is also shown for comparison.}
\label{fig:vqe}
\end{center}
\end{figure}

On our 5-qubit quantum computer (see Fig.~\ref{fig:vqe}) we aim to perform a VQE calculation in order to approximate the ground state of the one-impurity-site and one-bath-site AIM ($k = 1$ in Eq.~(\ref{eq:aim})). The fermionic Hamiltonian of the AIM has to be mapped to a qubit Hamiltonian using the following Jordan-Wigner transformation \cite{jordan_1928}:
\begin{align}
d_{\uparrow}^\dagger & = \sigma_0^- = \frac{1}{2} (X_0 - i Y_0) \\
c_{1\uparrow}^\dagger & = Z_0 \sigma_1^- = \frac{1}{2} Z_0 (X_1 - i Y_1) \\
d_{\downarrow}^\dagger & = Z_0 Z_1 \sigma_2^- = \frac{1}{2} Z_0 Z_1 (X_2 - i Y_2) \\
c_{1\downarrow}^\dagger & = Z_0 Z_1 Z_2 \sigma_3^- = \frac{1}{2} Z_0 Z_1 Z_2 (X_3 - i Y_3) 
\end{align}
which leads to the qubit Hamiltonian
\begin{align}
\mathcal{H}_{\text{AIM}}^{\text{JW}}  & = (\varepsilon_d + \varepsilon_1 - 2\mu) -  \frac{1}{2}(\varepsilon_d -\mu +2U)(Z_0 + Z_2) -  \frac{1}{2}\varepsilon_1(Z_1 + Z_3)  \nonumber \\
& +  \frac{1}{4} U Z_0Z_2 + \frac{1}{2} V_1 (X_0X_1 + Y_0Y_1 + X_2X_3 + Y_2Y_3).
\end{align}
Grouping Hamiltonian terms that have support on different qubits allows us to reduce the number of measurements performed. For example an $X_0X_1X_2X_3$ measurement on the Ansatz state can be used to compute both $X_0X_1$ and $X_2X_3$ terms of the Hamiltonian. A hardware efficient Ansatz as shown in Fig.~\ref{fig:vqe} (a) is chosen. Since the problem requires only four qubits, the best qubits based on fidelity are chosen from the hardware to perform the calculations and keeping $0^{th}$ qubit of the spin Hamiltonian fixed at the center of the star topology. Parameterized single qubit $R_y$ gates are introduced and a total of seven parameters are tuned by the L-BGFS-B classical optimizer from Scipy-optimizer package \cite{sciPy}. This is a gradient based optimizer where the derivative is computed using the parameter shift rule \cite{schuld_2019,mitarai_2018}. One has:
\begin{align}
    \frac{\partial E_{{\bm {\theta}}} }{\partial \theta_i} = \frac{E_{\theta_i + \frac{\pi}{2}} - E_{\theta_i - \frac{\pi}{2}}}{2}
\end{align}
where $E_{{\bm {\theta}}} = \langle \psi({\bm {\theta}}) | \mathcal{H}_{\text{AIM}}^{\text{JW}} | \psi(\bm \theta) \rangle$ with ${\bm {\theta}} = \{\theta_1,\theta_2 \cdots \}$.  Figure~\ref{fig:vqe} (b) shows the energy measurements for different iterations computed without and with read-out error mitigation at $\epsilon_d = \mu, \epsilon_1 = 0, V=1$ and $U = 2$. We have used 5000 shots for each measurement. The results can be further improved using advanced error mitigation strategies, such as zero-noise extrapolation (ZNE) \cite{endo_2018, kandala_2019,giurgica_2020} or probabilistic error cancellation (PEC) \cite{temme_2017,van_2023,gupta_2023}. The converged parameterized state can be used to compute quantities such as density matrix or the Green's function which are necessary components of some of the aforementioned embedding techniques.  

VQE has been proposed as a promising algorithm for near-term quantum computers to solve certain models from quantum chemistry and condensed matter physics. It is important to note though, that practical implementations of VQE are still limited by the capabilities of available quantum hardware, and there is active ongoing work on improving and refining these algorithms. Nevertheless, it serves as a gateway to explore and investigate advanced embedding techniques on the currently available quantum hardware.

\section{Summary}
\label{sec:discussion}

\par An on-site quantum computer can be utilized in education and research for quantum computing, quantum information and quantum theory. 
 We have demonstrated some of these topics with the 5-qubit superconducting {IQM Spark\textsuperscript{TM}} prototype in this paper. 
\par First, we presented the tools and programs used. Then, we showed how they can be used in education and research. Certain demonstrations, such as calibration and working with qutrits, are only possible with an on-site quantum computer. It is also used to explore the complex quantum realm and reproduce recent breakthroughs in mathematics, physics, and chemistry for a better understanding. 

\section{Discussion}

\par This physical on-site quantum computer is vital for making quantum computing accessible to more people and talents, teaching quantum concepts, and enhancing our understanding of quantum theory and computing. It can also be used for research, as we have shown in Section 5. A recent survey reveals that many research papers demonstrating “proofs of principle” used a quantum computer with five qubits of fewer \cite{bib:survey}. 
\par We anticipate that in the very near-term future, every leading university and research institute, wanting to stay competitive in quantum computing education and research, will have physical access to affordable on-site quantum computers, such as the IQM Spark\textsuperscript{TM}~\cite{bib:spark}.


\section*{Acknowledgments}

The work of the IQM technology team is acknowledged for having created the technology used here in the educational and research setting. We would like to thank Roberto Moretti and Michihisa Wakui for enlightening discussions. We are grateful to Andrew Guthrie for careful reading of the manuscript. 

\section*{Funding}

 The whole work is funded by IQM Quantum Computers.

\section*{Availability of data and materials}

 The codes employed in this paper are available from the corresponding author upon request.

\section*{Competing interests}

IQM Spark$^\mathrm{TM}$ is a product of IQM, the employer of all authors

\section*{Authors’ contributions}

All authors contributed equally in this paper.

\bibliography{biblio}
\end{document}